\documentclass[aps,prl,twocolumn,groupedaddress,longbibliography,nobibnotes]{revtex4-1}
\usepackage{amsmath}
\usepackage{amssymb}
\usepackage{mathrsfs}
\usepackage[pdftex,colorlinks=true,urlcolor=blue,linkcolor=blue,citecolor=blue,breaklinks=true]{hyperref}
\usepackage{graphicx}
\usepackage{float}
\usepackage{dsfont}
\usepackage{multibib}
\newcommand{\push}{\hspace{0.05cm}}
\newcommand{\pull}{\hspace{-0.05cm}}

\makeatletter
\def\l@subsection{%
 \l@@sections{section}{subsection}
}
\def\l@f@subsection{%
 \addpenalty{\@secpenalty}%
 \addvspace{0.5em plus\p@}%
}
\def\l@subsubsection#1#2{}
\makeatother

\begin{document}

\title{Long-Range Coherence and Multiple Steady States in a Lossy Qubit Array}

\author{Shovan Dutta}
\email[E-mail: ]{sd843@cam.ac.uk}
\author{Nigel R. Cooper}
\email[E-mail: ]{nrc25@cam.ac.uk}
\affiliation{T.C.M. Group, Cavendish Laboratory, University of Cambridge, JJ Thomson Avenue, Cambridge CB3 0HE, United Kingdom\looseness=-1}

\date{\today}

\begin{abstract}
We show that a simple experimental setting of a locally pumped and lossy array of two-level quantum systems can stabilize states with strong long-range coherence. Indeed, by explicit analytic construction, we show there is an extensive set of steady-state density operators, from minimally to maximally entangled, despite this being an interacting open many-body problem. Such nonequilibrium steady states arise from a hidden symmetry that stabilizes Bell pairs over arbitrarily long distances, with unique experimental signatures. We demonstrate a protocol by which one can selectively prepare these states using dissipation. Our findings are accessible in present-day experiments.
\end{abstract}

\maketitle

{\it Introduction.---}Coupling a quantum system to an environment typically results in a loss of coherence \cite{Zurek2006}, which is a major obstacle for quantum control and information processing \cite{Palma1996, Bennett2000, Nielsen2010}. However, a growing number of studies have shown that a well-designed coupling can also drive the system toward interesting and useful quantum states \cite{Verstraete2009, Schirmer2010, Diehl2011}. This is particularly promising in light of parallel advances in experimental platforms where one can engineer both the system Hamiltonian and the coupling \cite{Poyatos1996, Barreiro2011, Mueller2012, Houck2012, Georgescu2014}, offering novel out-of-equilibrium settings where interactions and dissipation compete \cite{Sieberer2016, Rotter2015}.

A rare phenomenon occurs when such an open quantum system has multiple stable states owing to a symmetry that gives a conserved quantum number \cite{Baumgartner2008, Albert2014}. Then the dynamics decouple into independent sectors \cite{Buca2012}, with the system retaining some memory of its initial state \cite{Albert2016}. Furthermore, one can show that information encoded in the steady-state manifold would be preserved unconditionally \cite{BlumeKohout2010}, and one could control transport by switching between the symmetry sectors \cite{Manzano2018}. So far, this kind of strong symmetry has been found only theoretically in symmetric networks \cite{Manzano2014, Thingna2016, Thinga2020} and in boundary-driven spin chains with nonstandard dissipation \cite{Buca2012, Ilievski2014}, without experimental realizations. They require special design even in noninteracting systems \cite{Thinga2020}.

In this Letter, we identify a prototypical setting of a simple lattice model with a routine bulk dissipation that possesses a surprising hidden symmetry, leading to multiple steady states with long-range coherence and nonlocal Bell pairs. The steady states can be selectively prepared and probed in existing setups. To illustrate our findings, we model hard-core bosons on a one-dimensional (1D) lattice, which is equivalent to an array of qubits or an (anisotropic) spin chain. Our conclusions extend more generally to a broad class of models of this type.

We consider hard-core bosons on a lattice with particle injection and loss at two sites. Such a local incoherent pump was used recently to prepare a Mott insulator of photons \cite{Ma2019}. When the source and sink are at the boundary, the system can be reduced to free fermions \cite{Prosen2008, Prosen2008a, Kos2017}. However, for dissipation in the bulk, such a reduction is not possible and the system is strongly interacting. We focus on the special case where the pump and loss both act on the center site. Beyond the obvious reflection parity, we find a dynamical symmetry that can be roughly interpreted as conserving a total ``charge'' of symmetrically located particle-hole Bell pairs. Consequently, the number of symmetry sectors grows linearly with the system size $L$, yielding an extensive degeneracy. We provide an exact solution for the steady-state manifold and show that it includes a maximally entangled sector with $(L-1)/2$ nonlocal Bell pairs. We demonstrate a procedure for preparing the system in any given sector. Subsequent dynamics within the sector converge to a unique steady state, which can be discerned by measuring single-particle or density correlations \cite{Filipp2009, Titchener2018, Bergschneider2019}. Additionally, in the limit of zero pump (or loss) rate, the degeneracy is increased further to accommodate a decoherence-free subspace, a key ingredient for quantum computing \cite{Lidar2003}.

{\it Model.---}We study hard-core bosons \cite{Cazalilla2011} hopping on a 1D lattice with an odd number of sites, $L:=2l+1$ for integer $l$, described by the Hamiltonian
\begin{equation}
\hat{H} = -\hbar J \sum\nolimits_{i=-l}^{l-1} \big(\hat{b}_{i}^{\dagger} \hat{b}_{i+1} + \hat{b}_{i+1}^{\dagger} \hat{b}_{i}\big)\;,
\label{bosonhamil}
\end{equation}
where $J$ is the hopping amplitude and $\smash{\hat{b}_i^{\dagger}}$ creates a boson at site $i$. The hard-core condition is imposed by requiring $\smash{\hat{b}_i^{\dagger 2}=0}$, which ensures there can be either 0 or 1 particle at any given site. This regime corresponds to the strong-interaction limit of the Bose-Hubbard model \cite{Kordas2015} and has been realized with atoms in optical lattices \cite{Paredes2004, Stoeferle2004, Preiss2015} and with photons in nonlinear resonators \cite{Ma2019}. The hard-core constraint implies the commutation rules $\smash{[\hat{b}_i,\hat{b}_j]=0}$ and $\smash{[\hat{b}_i,\hat{b}_j^{\dagger}] = (-1)^{\hat{n}_i} \delta_{ij}}$, where $\smash{\hat{n}_i:=\hat{b}_i^{\dagger}\hat{b}_i}$ is the occupation at site $i$ \cite{matsubara1956lattice}. The Hamiltonian maps onto free fermions by a Jordan-Wigner (JW) transformation \cite{Jordan1928}:
\begin{equation}
\hat{f}_j = (-1)^{\sum_{i<j} \hat{n}_i} \hat{b}_j\;;\quad \hat{b}_j = (-1)^{\sum_{i<j} \hat{n}_i} \hat{f}_j\;,
\label{JordanWigner}
\end{equation}
where $\hat{f}_j$ are fermionic operators that satisfy anticommutation, $\smash{\{\hat{f}_i,\hat{f}_j\}=0}$ and $\smash{\{\hat{f}_i,\hat{f}_j^{\dagger}\}=\delta_{ij}}$. Thus, $\smash{\hat{f}_i^{\dagger}\hat{f}_i = \hat{n}_i}$, and Eq.~\eqref{bosonhamil} is restated as $\hat{H} = -\hbar J\sum_i \big(\hat{f}_i^{\dagger} \hat{f}_{i+1} + \text{H.c.}\big)$. However, as we show below, the dissipation mediates interactions between these fermion operators.

We add dissipation by coupling the system of bosons to bosonic reservoirs that inject particles at site $p$ and remove particles from site $q$. The reservoirs have a finite bandwidth, such that if site $p$ is already occupied, further injection is suppressed by the large interaction energy. Such local sources and sinks have been engineered using transmon qubits in microwave circuits \cite{Ma2019}. Typically, in these photonic setups, the reservoirs relax to equilibrium much faster than the system dynamics \cite{Daley2014}. Under such a routine Born-Markov approximation, the reduced density matrix $\hat{\rho}$ of the system is governed by a master equation of the Lindblad form \cite{Lindblad1976, Gorini1976, carmichael2013statistical, Kordas2015, Daley2014, rivas2010markovian, dhahri2008lindblad, maghrebi2016nonequilibrium}
\begin{equation}
\frac{d\hat{\rho}}{dt} = \mathcal{L}\hat{\rho} := -\frac{i}{\hbar}\push [\hat{H},\hat{\rho}] \push + \sum\nolimits_{\alpha}\pull \hat{L}_{\alpha} \hat{\rho}\hat{L}_{\alpha}^{\dagger} - \frac{1}{2} \{\hat{L}_{\alpha}^{\dagger} \hat{L}_{\alpha}, \hat{\rho}\}\;,
\label{mastereqn}
\end{equation}
where $\smash{\hat{L}_1 := \sqrt{\gamma_+}\push\hat{b}_p^{\dagger}}$ and $\smash{\hat{L}_2 := \sqrt{\gamma_-}\push\hat{b}_q}$ are two Lindblad operators, $\gamma_{\pm}$ being the pump and loss rates, respectively. This dynamics could also be realized with cold atoms by mapping the system to a spin-1/2 XX chain with local incoherent spin flips~\cite{Cazalilla2011}. Such a chain could be engineered with either motional states \cite{schwager2013dissipative} or internal states \cite{duan2003controlling, Mamaev2020, Browaeys2020}, using local addressability to flip spins \cite{fukuhara2013quantum}. Note our main results do not depend on the exact form of the Lindblad operators as long as they are local.

If the pump and loss are at the ends of the chain (i.e., $|p|=|q|=l$), the problem reduces to a description in which the Liouvillian $\mathcal{L}$ is quadratic in the JW fermions and the system is noninteracting \cite{Prosen2008, Prosen2008a, Kos2017, weaksym}. If either pump or loss occurs in the bulk, this can no longer be achieved. Then $\mathcal{L}$ contains terms involving string operators $\smash{(-1)^{\hat{N}_L}}$, where $\hat{N}_L$ is the number of particles to the left of the dissipation site, which is not conserved by the Hamiltonian. Consequently, $\mathcal{L}$ is not quadratic and the system is genuinely interacting.

Here, we focus on these cases where pump or loss does not occur at the boundary. For such interacting systems, one expects that, under generic conditions, Eq.~\eqref{mastereqn} has a unique steady state \cite{Spohn1977, Evans1977, uniqueness}. We find this is indeed the case if the pump or loss occurs at any site other than the center \cite{dutta2020out}. For $p=q\neq 0$, the system reaches a product state $\smash{\hat{\rho} = \otimes_i (\gamma_+ |1_i\rangle \langle 1_i| \hspace{-0.01cm}+\hspace{-0.01cm} \gamma_- |0_i\rangle \langle 0_i|)/(\gamma_+ \pull + \gamma_-)}$ \cite{Pizorn2013}. The situation is very different, however, if the pump and loss are both at the center site, unlocking multiple ``strong" symmetries \cite{Buca2012} and leading to many striking effects.

{\it Hidden symmetry.---}To understand the symmetries that arise when both pump and loss occur at the center,  $p=q=0$, consider first the reflection symmetry. Reflections are generated by an operator   $\hat{R}$ that exchanges sites $i$ and $-i$ for all $i$, such that $\smash{\hat{R}\push\hat{b}_i\hat{R} = \hat{b}_{-i}}$. One can readily show that  $\hat{R}$ commutes with both the Hamiltonian and the dissipators:
\begin{equation}
[\hat{H},\hat{R}] = 0\quad\text{and}\quad [\hat{L}_{\alpha},\hat{R}]=0\;\;\forall \alpha\;,
\label{strongsymmetry}
\end{equation}
the latter arising since the dissipators involve only $\smash{\hat{b}^{(\dag)}_0}$ and $\smash{[\hat{b}^{(\dag)}_0, \hat{R}]=0}$. Consequently, reflection $\hat{R}$ generates a so-called  ``strong" symmetry \cite{Buca2012} and leads to multiple steady states. Here, the system evolves separately in its even- and odd-parity sectors, giving rise to (at least) two steady states associated with the two parities. 

The dynamics are far more constrained, however, by a hidden symmetry \cite{cariglia2014hidden} generated by another operator $\hat{C}^2$, where
\begin{equation}
\hat{C} := -1/2 + \sum\nolimits_{k=-l}^l \hat{f}_{k}^{\dagger}\hat{f}_{-k} \;.
\label{defineC}
\end{equation}
From Eq.~\eqref{JordanWigner}, every $k\neq 0$ term in $\hat{C}$ contains the factor $\smash{(-1)^{\hat{n}_0}}$, and the remaining terms give $\smash{\hat{n}_0\pull -1/2 \propto (-1)^{\hat{n}_0}}$. Thus, 
$\smash{[\hat{b}_0^{(\dag)},\hat{C}^2]=0}$; i.e., $\hat{C}^2$ commutes with the dissipators $\hat{L}_{1,2}$. Furthermore, as we show in the Supplemental Material (SM) \cite{supplement}, $\smash{\hat{C} = \hat{N}_{\text{even}} \pull-\pull \hat{N}_{\text{odd}} \pull-\pull 1/2}$, where $\hat{N}_{\text{even}}$ and $\hat{N}_{\text{odd}}$ are the total occupations of the even and odd single-particle energy modes, which gives $\smash{[\hat{H},\hat{C}]=0}$ \cite{quadraticsym}. Therefore, $\smash{\hat{C}^2}$ generates a strong symmetry. Note this is an exact result for the hard-core bosons. One also finds $\hat{C}$ is symmetric under reflection about the center, and all of its eigenspaces have a definite $\hat{R}$ parity.

In general, the eigenspaces of a strong symmetry generator $\hat{S}$ evolve independently, each having at least one steady state \cite{Buca2012}. This decoupling originates from conservation laws. In particular, using $\langle\hat{S}\rangle = \text{Tr}(\hat{\rho}\hat{S})$ in Eq.~\eqref{mastereqn}, one finds $d\langle\hat{S}\rangle/dt = i\langle[\hat{H},\hat{S}]\rangle/\hbar - \sum_{\alpha} \text{Re}\push \langle\hat{L}_{\alpha}^{\dagger} [\hat{L}_{\alpha},\hat{S}]\rangle=0$; i.e., $\smash{\langle\hat{S}\rangle}$ is conserved \cite{Gough2015, Albert2014}. Moreover, the projectors onto each of the eigenspaces of $\hat{S}$ satisfy Eq.~\eqref{strongsymmetry} individually and are conserved separately \cite{Baumgartner2008}. In other words, the weight in each symmetry sector is preserved.

Here, there will appear multiple steady states associated with the different eigenspaces of $\smash{\hat{C}^2}$. As we explain below, the eigenstates of $\smash{\hat{C}^2}$ comprise entangled particle-hole pairs at sites $k$ and $-k$, each carrying a quantum number taking values $\pm 1$ that we call ``charge.'' The full spectrum consists of $l+1$ eigenvalues, $\{(\eta+1/2)^2:\eta=0,\dots,l\}$, where $\eta$ is a measure of the total charge of all such pairs. These eigenspaces evolve independently, and we find every sector has a unique steady state for $\gamma_{\pm}\neq 0$, leading to an $(l+1)$-fold degeneracy. This is in sharp contrast to the noninteracting problem, where free bosons \cite{Kepesidis2012} or fermions are subject to pump or loss at the center. Then, every odd single-particle state is unaffected by dissipation so its occupation number is conserved, yielding an exponentially large decoherence-free subspace of degenerate steady states. Later, we will use this feature for preparing the symmetry sectors of $\hat{C}^2$.

{\it Steady states.---}We first characterize the eigenspaces of $\hat{C}$ which is written as a sum of $l+1$ commuting parts, $\smash{\hat{C}_0 := \hat{n}_0-1/2}$ and $\smash{\hat{C}_k :=\hat{f}_{k}^{\dagger}\hat{f}_{-k}+\text{H.c.}}$ for $k=1,\dots,l$. The latter describes hopping of JW fermions between two sites and can be diagonalized as $\smash{\hat{C}_k = \sum_{s=\pm} s\push \hat{a}_{k,s}^{\dagger} \hat{a}_{k,s}}$, where $\hat{a}_{k,\pm} := (\hat{f}_{k}\pm \hat{f}_{-k})/\sqrt{2}$ are single-particle fermion modes. Thus, $\hat{C}_k$ has eigenstates $\prod_{s=\pm} \pull \big(\hat{a}_{k,s}^{\dagger}\big)^{\nu_{k,s}} |0\rangle$ with eigenvalue $\nu_{k,+}\pull -\nu_{k,-}$, where $|0\rangle$ is the vacuum and $\nu_{k,\pm}\pull \in \{0,1\}$. One can think of $\smash{\hat{a}_{k,\pm}^{\dagger}}$ as creating a particle-hole pair of charge $\pm 1$ at sites $k$ and $-k$, of the form $(|01\rangle \pm |10\rangle)/\sqrt{2}$. The net charge is 0 for the states $|00\rangle$ and $|11\rangle$ \cite{spininterpretation}. It follows that the eigenstates of $\hat{C}$ are given by
\begin{equation}
|\{\nu_{k,\pm}\},n_0\rangle:=\big(\hat{f}_0^{\dagger}\big)^{\pull n_0} \prod_{k=1}^l \prod_{s=\pm} \big(\hat{a}_{k,s}^{\dagger}\big)^{\nu_{k,s}} \push |0\rangle\;,
\label{eigenC}
\end{equation}
with eigenvalue $\lambda = \nu+n_0 - 1/2$, where $\nu := \sum_k \nu_{k,+}\pull -\nu_{k,-}$. The integer $\nu$ measures the total charge of all Bell pairs and varies from $-l$ to $l$. Since $n_0$ is either 0 or 1, $\lambda$ can assume $2(l+1)$ distinct values, $\{\pm (\eta+1/2):\eta=0,\dots,l\}$ with degeneracies $\binom{L}{l-\eta}$.

\begin{figure}
\centering
\includegraphics[width=\columnwidth]{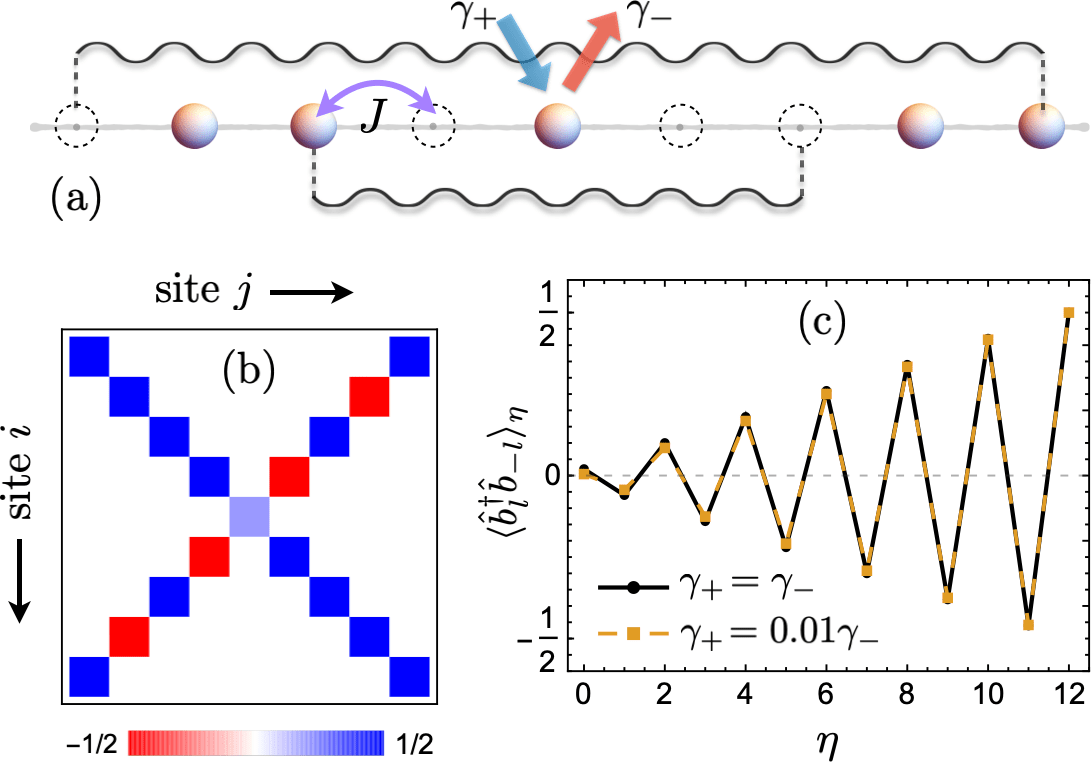}
\caption{\label{corrplots}(a) Schematic setup showing coherent tunneling $J$ and incoherent pump and loss $\gamma_{\pm}$ at the center. The system has a strong dynamical symmetry that stabilizes long-range entangled particle-hole pairs at reflection-symmetric sites. (b) Single-particle density matrix $\smash{\langle\hat{b}_i^{\dagger} \hat{b}_j\rangle}$ for the steady state with maximum number of Bell pairs. The center has occupation $\gamma_+/(\gamma_+ \pull + \gamma_-)$. (c) End-to-end coherence for different symmetry sectors $\eta$.
}
\end{figure}

The eigenstates in Eq.~\eqref{eigenC} share some general features which will be inherited by the steady states. In particular, using $\langle\hat{f}_{k}^{\dagger}\hat{f}_{-k}\rangle = \langle \hat{C}_k\rangle/2$ and transforming back to bosons, one finds they have an antidiagonal string order with long-range coherences, $\smash{|\langle \hat{b}_k^{\dagger}\hat{b}_{-k}\rangle|=|\nu_{k,+} \pull -\nu_{k,-}|/2}$, as illustrated in Fig. \ref{corrplots}(a). The sectors labeled by $\eta=l$ are nondegenerate and maximally entangled, containing $l$ Bell pairs of the same charge, with $\smash{\langle \hat{b}_k^{\dagger}\hat{b}_{-k}\rangle=(-1)^k/2}$ [Fig.~\ref{corrplots}(b)]. It can also be shown that the reflection parity is even if $\eta$ is of the form $4m$ or $4m+3$ for integer $m$ and odd otherwise (see SM \cite{supplement}). The same eigenstates diagonalize $\hat{C}^2$ with eigenvalue $\lambda^2=(\eta+1/2)^2$, generating $l+1$ distinct symmetry sectors.

To find the steady states in each sector, we define $\smash{\hat{P}_{\eta}}$ as the projector onto the corresponding eigenspace, $\hat{N}$ as the total particle number, and $\smash{\hat{P}_{\eta}^{\prime} :=(\gamma_+/\gamma_-)^{\hat{N}} \hat{P}_{\eta}}$. Note that $\smash{[\hat{H},\hat{P}_{\eta}^{\prime}]=0}$, as $\hat{H}$ commutes with both $\hat{N}$ and $\smash{\hat{P}_{\eta}}$. Further, since $\smash{\hat{C}^2}$ does not act on the center site, one has the form $\smash{\hat{P}_{\eta}^{\prime}=\hat{Q}_{\eta}\otimes (\gamma_+ |1\rangle\langle 1| + \gamma_-|0\rangle\langle 0|)}$, where $|0\rangle$ and $|1\rangle$ describe the center site and $\smash{\hat{Q}_{\eta}}$ acts on the remaining sites. These two properties imply that $\smash{\hat{\rho}_{\eta}:=\hat{P}_{\eta}^{\prime}/\text{Tr}(\hat{P}_{\eta}^{\prime})}$ is a steady state of Eq.~\eqref{mastereqn} with the dissipators $\smash{\sqrt{\gamma_+}\push\hat{b}_0^{\dagger}}$ and $\smash{\sqrt{\gamma_-}\push\hat{b}_0}$. Numerically, we find this is the only steady state in each sector \cite{Popkov2020}, up to the largest systems tractable by exact diagonalization. Within the respective eigenspace, $\smash{\hat{\rho}_{\eta}}$ describes an infinite-temperature state with chemical potential $\mu\pull =\ln(\gamma_+/\gamma_-\pull)$. Note, however, that such a state can have high spatial entanglement, as we discuss below. For numerics, we compute $\smash{\hat{\rho}_{\eta}}$ by generating all eigenstates of $\smash{\hat{C}^2}$ by repeated applications of $\smash{\hat{a}_{k,\pm}^{\dagger}}$ [Eq.~\eqref{eigenC}] and then forming $\smash{\hat{P}_{\eta}^{\prime}}$. A general steady state is given by $\smash{\hat{\rho}_{\infty} = \sum_{\eta} w_{\eta} \hat{\rho}_{\eta}}$ with $\sum_{\eta}\pull w_{\eta}=1$, where $w_{\eta}\geq 0$, since $\hat{\rho}$ must be positive semidefinite. The coefficients $w_{\eta}$ can be identified as the weights $\smash{\langle\hat{P}_{\eta}\rangle}$ in different symmetry sectors, which are the constants of motion. This gives a mapping from an initial state, characterized by $\smash{\langle\hat{P}_{\eta}\rangle}$, to the final state \cite{Munoz2019}:
\begin{equation}
\hat{\rho}_{\infty} = \sum_{\eta=0}^{l}\push \langle\hat{P}_{\eta}\rangle \push \frac{(\gamma_+/\gamma_-\pull )^{\hat{N}} \hat{P}_{\eta}}{\text{Tr}\big[(\gamma_+/\gamma_-\pull )^{\hat{N}} \hat{P}_{\eta}\big]} \;.
\label{steadystate}
\end{equation}
Note that $\hat{\rho}_{\infty}$ is fully determined by the weights $\smash{\langle\hat{P}_{\eta}\rangle}$ and the pump-to-loss ratio, irrespective of the tunneling $J$.

{\it Properties.---}The steady states $\hat{\rho}_{\eta}$ have unique signatures in the one-particle correlations $\smash{\langle\hat{b}_{k}^{\dagger} \hat{b}_{-k}\rangle}$, which can be measured experimentally \cite{Filipp2009, Titchener2018, Bergschneider2019} and have closed-form analytic expressions derived in the SM \cite{supplement}. In particular, the end-to-end coherence grows steadily with $\eta$ (in magnitude), $\smash{\langle\hat{b}_{l}^{\dagger} \hat{b}_{-l}\rangle = (-1)^{\eta} \push (\eta +1/2)/L}$ for $\gamma_+\pull = \gamma_-$, with a weak dependence on $\gamma_+/\gamma_-$, as shown in Fig.~\ref{corrplots}(c). Similar signatures appear in the density-density correlations (see SM \cite{supplement}). Note the correlations are symmetric under the exchange $\gamma_+ \leftrightarrow \gamma_-$. The occupations are $\langle\hat{n}_{k}\rangle=1/2$ for $\gamma_+=\gamma_-$ and grow monotonically with $\gamma_+/\gamma_-$, such that $\langle\hat{n}_{k\neq 0}\rangle = [1 \pm (1-\eta/l)]/2$ for $\gamma_{\mp}\pull\to 0$.

\begin{figure}
\centering
\includegraphics[width=\columnwidth]{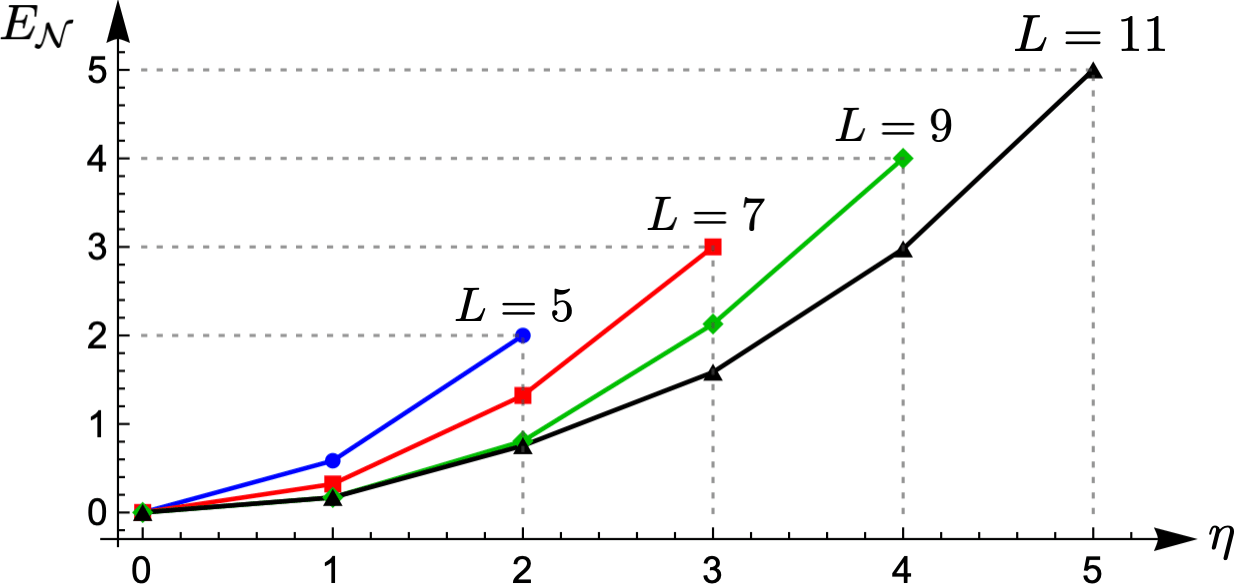}
\caption{\label{negativityplot}Log negativity $E_{\mathcal{N}}$, measuring entanglement between left and right halves, in steady states corresponding to different symmetry sectors $\eta$ and number of sites $L$, with $\gamma_{+} \pull=\gamma_-$.}
\end{figure}

To quantify the degree of entanglement in the (mixed) steady states, we numerically compute the log negativity $\smash{E_{\mathcal{N}}}$, which gives an upper bound on the number of distillable Bell pairs between two halves of the system~\cite{Vidal2002, Plenio2014, Horodecki2009}. As shown in Fig.~\ref{negativityplot}, it rises monotonically from $\smash{E_{\mathcal{N}}=0}$ for $\eta=0$ to $\smash{E_{\mathcal{N}}}=l$ for $\eta=l$. Other measures of coherence \cite{Streltsov2017} give similar results (see SM \cite{supplement}).

{\it Experimental preparation.---}Preparing this system of hard-core bosons in different symmetry sectors requires a controlled generation of entanglement. As we now show, this can be done by dissipative means \cite{Plenio2002, Kraus2008, Krauter2011, Ticozzi2012} if one can engineer loss of the JW fermions from the center site. For bosonic systems, such a process necessitates the application of a string operator, $\smash{\hat{f}_0 = \big[\prod_{i<0}(-1)^{\hat{n}_i}\big] \hat{b}_0}$, i.e., a boson loss accompanied by a collective phase. This can be realized efficiently in hardware with superconducting qubits coupled to ancilla cavities, as detailed in Ref.~\cite{Zhu2018}. For a spin realization with cold atoms, it would be more challenging but could be implemented, in principle, with a projective measurement of the spin coupled with local Zeeman fields. 

We target states in each sector that are made up solely of negatively charged Bell pairs, $\smash{\prod_{k=1}^l \pull \big(\hat{a}_{k,-}^{\dagger}\big)^{\nu_{k,-}} |0\rangle}$ with $\smash{\sum_{k} \nu_{k,-}\pull =\eta}$. Such states span the space of odd fermionic wavefunctions with occupation $N_{\text{odd}} \pull= \eta$, i.e., a total of $\eta$ JW fermions occupying the odd modes, which are linear combinations of $\{\smash{\hat{a}_{k,-}^{\dagger}}\}$ [Recall, $\smash{\hat{C} = \hat{N}_{\text{even}} \pull-\pull \hat{N}_{\text{odd}} \pull-\pull 1/2}$]. These modes are stable if one only has loss of the (now free) JW fermions at the center site. The same loss can be used to {\it produce} odd states with a given particle number, as follows. We start from a symmetric Fock state, $\smash{|\{n_k\}\rangle := \prod_{k=1}^l \big(\hat{b}_k^{\dagger} \hat{b}_{-k}^{\dagger}\big)^{\pull n_k} \big(\hat{b}_0^{\dagger}\big)^{\pull n_0} |0\rangle}$. Transforming to JW fermions, one finds such a state has $\smash{N_{\text{odd}}\hspace{-0.03cm}=\sum_{k=1}^l n_k}$ (see SM \cite{supplement}). Under JW fermion loss, only the even modes are depleted, so with the odd ones preserved the system will be driven to the sector $\smash{\eta=\sum_{k} n_k}$. Thus, one can selectively prepare all different sectors simply by setting the initial occupations. In particular, a fully filled lattice evolves to the maximally entangled state, $\eta=l$.

\begin{figure}
\centering
\includegraphics[width=\columnwidth]{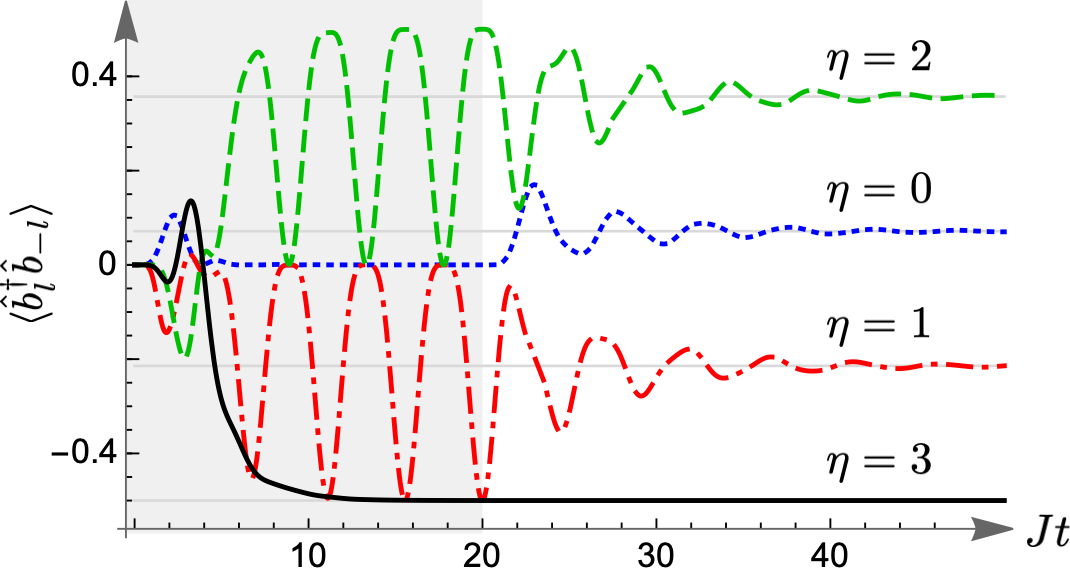}
\caption{\label{prepplot}End-to-end correlation during a selective preparation of different symmetry sectors $\eta$ for $L=7$. The shaded region shows the generation of Bell pairs from symmetric Fock states, $\smash{\hat{b}_0^{\dagger}  \prod_{k=1}^{\eta} \hat{b}_k^{\dagger} \hat{b}_{-k}^{\dagger} |0\rangle}$, driven by the loss of JW fermions at the center with rate $\smash{\gamma_F = 3J}$. The white region shows subsequent dynamics with boson pump and loss rates $\gamma_{\pm}=2J$.}
\end{figure}

Simulations of this preparation scheme, using exact diagonalization, are shown in Fig.~\ref{prepplot}. The oscillations describe breathing-type back-and-forth motion of the Bell pairs under the Hamiltonian. Once a given sector is prepared, one can switch from the JW fermion loss to the original boson pump and loss, converging to the steady state $\smash{\hat{\rho}_{\eta}}$. Note the preparation takes a few tens of tunneling time, much faster than the on-site disorder and residual dissipation in a recent experiment \cite{Ma2019}. We analyze these timescales further in SM \cite{supplement}.

{\it No-pump (or no-loss) limit.---}If one has only boson loss at the center ($\gamma_+=0$), all the odd JW fermionic modes become immune to dissipation. This is because they have no particle at the center site and are eigenmodes of $\hat{H}$. Hence, any superposition of these modes evolves unitarily under $\hat{H}$, yielding a decoherence-free subspace \cite{Lidar2003}. The same is also true for the loss of JW fermions at the center. However, the full dynamics are very different for the two cases \cite{Malo2018}. For the boson loss, $\smash{\langle\hat{C}^2\rangle}$ is conserved, not the occupation of odd modes, as the string in Eq.~\eqref{JordanWigner} couples odd and even modes. Thus, an initially filled lattice approaches the vacuum at long times, not the maximally entangled state in Fig.~\ref{prepplot}. With boson pump instead of loss, the particle and hole states are interchanged, and one again finds a decoherence-free subspace.

{\it Robustness.---}The observable $\smash{\hat{C}^2}$ remains a generator of strong symmetry for a large class of 1D systems. First, it is unaffected by dephasing \cite{carmichael2013statistical} or any Lindbladian dissipation at the center site. Second, as we show in the SM \cite{supplement}, $\hat{C}$ commutes with any Hamiltonian that is quadratic in the JW fermions and reflection symmetric, $\smash{[\hat{H},\hat{R}]=0}$. This includes symmetric trapping potentials, anisotropic XY spin-1/2 chains~\cite{Prosen2008a}, and the quantum Ising model which maps onto the Kitaev chain \cite{fendley2012parafermionic}. Third, the results are unaltered for periodic boundary conditions (with odd number of sites; see SM \cite{supplement} for more details). Furthermore, since $\smash{[\hat{C},\hat{n}_j+\hat{n}_{-j}]=0}$ for all $j$, any interactions or dissipation that depend only on such ``pair occupations" commute with $\hat{C}$. The symmetry is, however, broken for generic dissipation away from the center and for nearest-neighbor interactions of the form $\hat{H}^{\prime} = \varepsilon\sum_j \hat{n}_j \hat{n}_{j+1}$ (see SM \cite{supplement}), as found by mapping an XXZ chain to hard-core bosons. In such cases, the steady states are robust to linear order in Hamiltonian perturbations $\varepsilon$ \cite{tindall2020quantum, supplement}

{\it Conclusions.---}We have identified a paradigmatic experimental setting of a qubit array with local dissipation that exhibits a striking hidden symmetry, leading to stable long-range coherence that is both unusual and desirable. The symmetry stabilizes Bell pairs over arbitrarily long distances and is surprisingly robust. Consequently, the system has an extensive set of exactly solvable steady states characterized by an antidiagonal string order, from minimally to maximally entangled. We have shown how one can selectively prepare these states using dissipation, and discern them by correlation measurements, accessible in existing photonic \cite{Ma2019} and atomic setups. The controllable generation and preservation of long-range entanglement in an open platform would be valuable for quantum information processing and metrology \cite{Streltsov2017, BlumeKohout2010, Horodecki2009, Schindler2013, Huang2015}. Our findings of these special features in a simple paradigmatic model strongly motivate experimental investigations of symmetry in open systems, shedding light on the subtle relation between symmetry and conservation laws in a nonunitary setting \cite{Baumgartner2008, Albert2014}.

We thank Fabian Essler, Jon Simon, Dave Schuster, and Berislav Bu{\v{c}}a for valuable discussions. This work was supported by the Engineering and Physical Sciences Research Council Grant No.~EP/P009565/1 and by a Simons Investigator award.

\begingroup
\renewcommand{\addcontentsline}[3]{}
\renewcommand{\section}[2]{}


%

\endgroup


\onecolumngrid
\clearpage

\begin{center}
\textbf{\large Supplemental Material for\\``Long-range coherence and multiple steady states in a lossy qubit array"}
\end{center}

\setcounter{equation}{0}
\setcounter{figure}{0}
\setcounter{table}{0}
\setcounter{page}{1}
\setcounter{secnumdepth}{3}
\makeatletter

\renewcommand{\thefigure}{S\arabic{figure}}
\renewcommand{\theequation}{S\arabic{equation}}
\renewcommand{\bibnumfmt}[1]{[S#1]}
\renewcommand{\citenumfont}[1]{S#1}
\renewcommand{\thesection}{S\Roman{section}}

\renewcommand{\theHfigure}{S\thefigure}

\tableofcontents

\section{\label{hiddensym}Characterizations of the hidden symmetry}
In this section we derive the principal features of the hidden symmetry operator discussed in the main article. As before, we consider hard-core bosons on a 1D lattice, equivalent to an array of qubits or a spin-1/2 XX chain, described by the Hamiltonian
\begin{equation}
\hat{H} = -\hbar J \sum_{i=-l}^{l-1} \big(\hat{b}_{i}^{\dagger} \hat{b}_{i+1} + \hat{b}_{i+1}^{\dagger} \hat{b}_{i}\big)\;,
\label{sbosonhamil}
\end{equation}
where the boson operators satisfy $\smash{[\hat{b}_i,\hat{b}_j]=0}$ and $\smash{[\hat{b}_i,\hat{b}_j^{\dagger}] = (-1)^{\hat{n}_i} \delta_{ij}}$, with occupation $n_i \in \{0,1\}$. The Hamiltonian is mapped onto free fermions by the Jordan-Wigner (JW) transformation
\begin{equation}
\hat{f}_j = (-1)^{\sum_{i<j} \hat{n}_i} \hat{b}_j\;;\quad \hat{b}_j = (-1)^{\sum_{i<j} \hat{n}_i} \hat{f}_j\;,
\label{sJordanWigner}
\end{equation}
where $\smash{\{\hat{f}_i,\hat{f}_j\}=0}$ and $\smash{\{\hat{f}_i,\hat{f}_j^{\dagger}\}=\delta_{ij}}$, yielding
\begin{equation}
\hat{H} = -\hbar J \sum_{i=-l}^{l-1} \big(\hat{f}_{i}^{\dagger} \hat{f}_{i+1} + \hat{f}_{i+1}^{\dagger} \hat{f}_{i}\big)\;.
\label{sfermionhamil}
\end{equation}
The hard-core bosons are subject to pump and loss at the center site with rates $\gamma_{\pm}$, leading to non-unitary dynamics modeled by two Lindblad operators $\smash{L_1 = \sqrt{\gamma_+} \push \hat{b}_0^{\dagger}}$ and $\smash{L_2 = \sqrt{\gamma_-} \push \hat{b}_0}$, respectively. As described in the main text, the system has multiple steady states with long-range coherence, which originate from a hidden symmetry. The symmetry is a generated by a Hermitian operator $\hat{C}^2$, where
\begin{equation}
\hat{C} := -1/2 + \pull\sum_{k=-l}^l \hat{f}_{k}^{\dagger}\hat{f}_{-k} \;.
\label{sdefineC}
\end{equation}
Below we characterize the salient properties of this operator that are relevant to understanding the long-time dynamics.

\subsection{\label{uniqueness}Strong symmetry and uniqueness}
The multiple steady states arise because $\smash{\hat{C}^2}$ generates a strong symmetry, i.e., it commutes with both the Lindblad operators $\smash{\hat{L}_{1,2}}$ and the Hamiltonian $\smash{\hat{H}}$ \cite{sBuca2012}. The former condition is satisfied only if an operator does not affect the center site, i.e., it is of the form $\smash{\hat{O} \otimes \hat{\mathds{1}}}$, where the identity acts on the center and $\smash{\hat{O}}$ is a function of $\smash{\{\hat{b}_j,\hat{b}_j^{\dagger}:j\neq 0\}}$. This is true for $\smash{\hat{C}^2}$ because $\smash{\hat{C}}$ can be expressed, using Eq.~\eqref{sJordanWigner}, as
\begin{equation}
\hat{C} = \bigg[\pull -\pull\frac{1}{2} + \push\sum_{k=1}^l \big(\hat{b}_k^{\dagger} \hat{b}_{-k} + \hat{b}_{-k}^{\dagger} \hat{b}_k\big) \pull\prod_{0<|i| <k} (-1)^{\hat{n}_i}\bigg]\otimes (-1)^{\hat{n}_0}\push .
\label{sCboson}
\end{equation}
In addition, $\smash{\hat{C}^2}$ is a symmetry of the Hamiltonian as $\smash{\hat{C}}$ itself commutes with $\hat{H}$. This is not immediately apparent, although note that $\hat{C}$ is a quadratic form in the fermion operators, and thus a natural candidate for symmetry of a free-fermion Hamiltonian. Below we show that $\smash{\hat{C}}$ is, in fact, the only quadratic form that commutes with $\hat{H}$ and has the structure $\smash{\hat{C} = \hat{O} \otimes(-1)^{\hat{n}_0}}$\pull. The same arguments can be used to show there is no such operator of the form $\smash{\hat{O} \otimes \hat{\mathds{1}}}$, which can generate a strong symmetry itself.

From Eq.~\eqref{sJordanWigner}, the most general (Hermitian) quadratic form in $\smash{\{\hat{f}_j,\hat{f}_j^{\dagger}\}}$ that is proportional to $(-1)^{\hat{n}_0}$ is given by
\begin{equation}
\hat{A} = \zeta\push (-1)^{\hat{n}_0} + \sum_{i,j>0} \alpha_{i,j} \hat{f}_i^{\dagger} \hat{f}_{-j} + \beta_{i,j} \hat{f}_i \hat{f}_{-j} + \text{H.c.}\push ,
\label{squadform}
\end{equation}
where $\zeta$, $\alpha_{i,j}$, and $\beta_{i,j}$ are arbitrary coefficients. To find $\smash{[\hat{H},\hat{A}]}$, we use Eq.~\eqref{sfermionhamil} and the identities $\smash{(-1)^{\hat{n}_0} = 1 - 2\hat{f}_0^{\dagger}\hat{f}_0}$ and $[ab,cd] = a\{b,c\}d + ca\{b,d\} - \{a,c\}bd - c\{a,d\}b$, yielding
\begin{align}
\nonumber \hspace{-0.2cm} (\hbar J)^{-1} [\hat{H},\hat{A}] &= \sum_{i>0} \big(\alpha_{i,1} + 2\zeta\delta_{i,1}\big) \hat{f}_i^{\dagger} \hat{f}_0 - \big(\alpha_{1,i} + 2\zeta\delta_{i,1}\big) \hat{f}_0^{\dagger} \hat{f}_{-i} + \beta_{i,1} \hat{f}_i\hat{f}_0 +\beta_{1,i} \hat{f}_0\hat{f}_{-i}\\
&+ \sum_{i,j>0} \big(\alpha_{i,j+1} + \alpha_{i,j-1} - \alpha_{i+1,j} - \alpha_{i-1,j}\big) \hat{f}_i^{\dagger} \hat{f}_{-j} + \big(\beta_{i,j+1} + \beta_{i,j-1} + \beta_{i+1,j} + \beta_{i-1,j}\big) \hat{f}_i \hat{f}_{-j} - \text{H.c.}\push,
\label{squadcom}
\end{align}
with the understanding that $\alpha_{i,j}$ and $\beta_{i,j}$ vanish outside the range $0<i,j\leq l$. For $\smash{[\hat{H},\hat{A}]}=0$, all the coefficients in Eq.~\eqref{squadcom} must vanish, which implies $\alpha_{i,j} = -2\zeta \delta_{i,j}$ and $\beta_{i,j}=0$ $\forall i,j$. The operator $\smash{\hat{C}}$ in Eq.~\eqref{sCboson} corresponds to the choice $\zeta=-1/2$. Thus, it is the only operator of the form in Eq.~\eqref{squadform} that commutes with $\smash{\hat{H}}$. An alternative derivation of $\smash{[\hat{H},\hat{C}]=0}$ is found by writing $\smash{\hat{C}}$ in terms of occupations of the energy eigenmodes (see Sec.~\ref{spmodes}).

\subsection{\label{robustness}Robustness}
Here we show that $\smash{\hat{C}}$ commutes with any reflection-symmetric Hamiltonian that is quadratic in the JW fermions. Further, we show this no longer holds in the presence of generic nearest-neighbor interactions. While the former class of Hamiltonians are generically nonlocal in the boson basis, they include symmetric potentials, anisotropic XY spin chains, as well as the transverse-field Ising model which transforms onto the Kitaev chain \cite{sfendley2012parafermionic}.

First, we consider general quadratic Hamiltonians of the form
\begin{equation}
\hat{H}_q = \sum_{i,j} \alpha_{i,j} \hat{f}_i^{\dagger} \hat{f}_j + \beta_{i,j} \hat{f}_i \hat{f}_j + \text{H.c.}\push,
\label{squadH}
\end{equation}
where $\alpha_{i,j}$ and $\beta_{i,j}$ are arbitrary coefficients. Such a Hamiltonian incorporates on-site potentials (for $i=j$), and all possible tunneling and pairing of the JW fermions. To see which of these Hamiltonians are symmetric under reflection about the center, we define a reflection operator $\smash{\hat{R}}$ which gives $\smash{\hat{R} \hat{b}_i \hat{R} = \hat{b}_{-i}}$ $\forall i$, as in the main text. Using Eq.~\eqref{sJordanWigner}, one finds the fermion operators transform as
\begin{equation}
\hat{R}\hat{f}_i\hat{R} = \hat{P} \hat{f}_{-i}\push,\quad\text{where}\quad \hat{P} := (-1)^{\hat{N}}
\label{sRfermion}
\end{equation}
is the total particle-number parity. It is also easy to show that $\smash{\{\hat{P},\hat{f}_i^{(\dagger)}\}=0}$. Thus, the Hamiltonian transforms as
\begin{equation}
\hat{R} \hat{H}_q \hat{R} = \sum_{i,j} \alpha_{i,j} \hat{f}_{-i}^{\dagger} \hat{f}_{-j} - \beta_{i,j} \hat{f}_{-i} \hat{f}_{-j} + \text{H.c.}\push .
\label{sRquadH}
\end{equation}
Comparing with Eq.~\eqref{squadH}, we find reflection symmetry implies $\alpha_{-i,-j} = \alpha_{i,j}$ and $\beta_{-i,-j} = -\beta_{i,j}$ $\forall i,j$, which makes intuitive sense. Now we show the same condition is derived by requiring $\smash{[\hat{H}_q,\hat{C}]=0}$. Using Eqs.~\eqref{sdefineC} and \eqref{squadH} yields
\begin{equation}
[\hat{H}_q,\hat{C}] = \sum_{i,j} \big(\alpha_{i,-j} - \alpha_{-i,j}\big) \hat{f}_i^{\dagger} \hat{f}_j + \big(\beta_{i,-j} + \beta_{-i,j}\big) \hat{f}_i \hat{f}_j - \text{H.c.},
\label{squadHC}
\end{equation}
which vanishes if and only if $\alpha_{i,-j} = \alpha_{-i,j}$ and $\beta_{i,-j} +\beta_{-i,j}=0$ $\forall i,j$, i.e., $\smash{\hat{H}_q}$ is reflection symmetric. This means the strong symmetry generated by $\smash{\hat{C}^2}$ is robust under any such reflection-symmetric Hamiltonian. Similarly, one can show that $\smash{[\hat{C},\hat{n}_j + \hat{n}_{-j}]=0}$ $\forall j$, so the strong symmetry will survive interactions or dissipation which are functions of $\hat{n}_j + \hat{n}_{-j}$.

Next we show the strong symmetry is broken in the presence of generic nearest-neighbor interactions of the form $\smash{\hat{H}_{\text{int}} = \sum_j\alpha_j \hat{n}_j \hat{n}_{j+1}}$, where $\alpha_j$ are the interaction strengths. Using Eq.~\eqref{sdefineC}, we find
\begin{equation}
[\hat{H}_{\text{int}},\hat{C}] = \sum_{j=1}^l \hat{f}_j^{\dagger} \hat{f}_{-j} \big[\alpha_{j-1}\push \hat{n}_{j-1} - \alpha_{-j}\push \hat{n}_{1-j} + (1-\delta_{j,l}) \big(\alpha_j\push \hat{n}_{j+1} - \alpha_{-j-1} \push\hat{n}_{-j-1}\big)\big] - \text{H.c.}\push,
\label{sHintC}
\end{equation}
which vanishes only if $\alpha_{-1} = \alpha_0$ and $\alpha_j = 0$ for all other $j$. Therefore, the only such Hamiltonian that commutes with $\smash{\hat{C}}$ is $\smash{\hat{H}_{\text{int}} = \alpha_0 \hat{n}_0(\hat{n}_{1}+\hat{n}_{-1})}$, which is a special case of the symmetry-preserving interactions of the form $\smash{(\hat{n}_j + \hat{n}_{-j})(\hat{n}_{j+1} + \hat{n}_{-j-1})}$. Similarly, one can show the symmetry is broken for generic pump or loss at any site other than the center. For a general symmetry-breaking Hamiltonian perturbation $\varepsilon \hat{H}^{\prime}$, one can show the steady states will be robust to first order in $\varepsilon$. This can be understood by starting from a steady state $\smash{\hat{\rho}_{\eta}}$ in a given eigenvalue sector of $\smash{\hat{C}^2}$ [Eq.~\eqref{ssteadystates}], and calculating
\begin{equation}
\langle \hat{C}^2 \rangle (t) - \langle \hat{C}^2 \rangle (0) 
= -i \varepsilon t  \;\text{Tr}\{\hat{C}^2 [\hat{H}^{\prime}, \hat{\rho}_{\eta}]\} + \mathcal{O}(\varepsilon^2 t^2)  
= i \varepsilon t \;\text{Tr}\{[\hat{C}^2, \hat{\rho}_{\eta}], \hat{H}^{\prime}\} + \mathcal{O}(\varepsilon^2 t^2)
= \mathcal{O}(\varepsilon^2 t^2)\;.
\label{spert}
\end{equation} 

\subsection{\label{stringorder}Eigenstates and string order}
Here we review the eigenstates of $\smash{\hat{C}}$ and calculate the associated single-particle correlations, which exhibit a long-range string order. As described in the main text, $\smash{\hat{C}}$ can be diagonalized as
\begin{equation}
\hat{C} = \hat{n}_0 - 1/2 + \sum_{k=1}^l \sum_{s=\pm} s \push \hat{a}_{k,s}^{\dagger} \hat{a}_{k,s}\push,
\label{sCdiag}
\end{equation}
where the eigenmodes $\smash{\hat{a}_{k,\pm}}$ are given by
\begin{equation}
\hat{a}_{k,\pm} := \frac{1}{\sqrt{2}} (\hat{f}_k \pm \hat{f}_{-k}), \quad k=1,\dots,l\push,
\label{seigenmodes}
\end{equation}
which satisfy fermionic anticommutation, $\smash{\{\hat{a}_{k,s},\hat{a}_{k' \pull,s'}\pull\}=0}$ and $\smash{\{\hat{a}_{k,s},\hat{a}_{k' \pull,s'}^{\dagger}\pull\}=\delta_{k,k'}} \delta_{s,s'}$. One can interpret the eigenmodes as describing entangled particle-hole pairs of ``charge'' $\pm 1$ at sites $k$ and $-k$. The eigenstates of $\smash{\hat{C}}$ are found by filling up these modes with occupation 0 or 1, yielding
\begin{equation}
|\{\nu_{k,\pm}\},n_0\rangle:=\big(\hat{f}_0^{\dagger}\big)^{\pull n_0} \prod_{k=1}^l \prod_{s=\pm} \big(\hat{a}_{k,s}^{\dagger}\big)^{\nu_{k,s}} \push |0\rangle\;,
\label{seigenC}
\end{equation}
with eigenvalue $\lambda = \nu + n_0 - 1/2$, where $\nu:=\sum_{k=1}^l (\nu_{k,+} - \nu_{k,-}\pull )$ and $\nu_{k,\pm} \in \{0,1\}$. Note the eigenstates have definite ``pair occupations,'' $n_k + n_{-k} = \nu_{k,+} + \nu_{k,-}$. This is because $\smash{\hat{C}}$ commutes with $\hat{n}_k + \hat{n}_{-k}$. Hence, the single-particle corrrelation $\smash{\langle\hat{b}_i^{\dagger} \hat{b}_j \rangle}$ is nonzero only if $j=\pm i$. One can see that the average occupations are given by
\begin{equation}
\langle\hat{b}_k^{\dagger} \hat{b}_k \rangle = \langle\hat{b}_{-k}^{\dagger} \hat{b}_{-k} \rangle = (\nu_{k,+} \pull + \nu_{k,-})/2 \;,
\label{snkeigen}
\end{equation}
for $k\geq 1$. To find the antidiagonal correlations, we use the JW transformation [Eq.~\eqref{sJordanWigner}] to write
\begin{equation}
\langle \hat{b}_k^{\dagger} \hat{b}_{-k} \rangle = \left\langle \hat{f}_k^{\dagger} \hat{f}_{-k} \push (-1)^{\sum_{|i|<k} \hat{n}_i}  \right\rangle = (-1)^{n_0 + \sum_{i=1}^{k-1} (\nu_{i,+} + \nu_{i,-})} \left\langle \hat{f}_k^{\dagger} \hat{f}_{-k} \right\rangle \push.
\label{santidiagconvert}
\end{equation}
Then using Eq.~\eqref{seigenmodes} and the anticommutation of the eigenmodes yields
\begin{equation}
\langle \hat{b}_k^{\dagger} \hat{b}_{-k} \rangle = \frac{1}{2} (\nu_{k,+} - \nu_{k,-})\; (-1)^{n_0 + \sum_{i=1}^{k-1} (\nu_{i,+} + \nu_{i,-})} \push .
\label{sstringorder}
\end{equation}
Thus, we find a string order that depends on the number of particles between sites $k$ and $-k$. In particular, for states that are composed of $l$ Bell pairs of the same charge, one finds $\smash{\langle \hat{b}_k^{\dagger} \hat{b}_{-k} \rangle = \pm (-1)^{n_0+k}/2}$, where the $+$ and $-$ signs correspond to negative and positive charges, respectively. These are the maximally entangled states.

\subsection{\label{parity}Reflection parity}
Here we show the eigenstates of $\smash{\hat{C}}$ in Eq.~\eqref{seigenC} all have definite reflection parity that depends only on the eigenvalue $\lambda$. First, we note that $\smash{\hat{C}}$ commutes with the reflection operator $\smash{\hat{R}}$ defined in Sec.~\ref{robustness}. This can be seen by using the transformation in Eq.~\eqref{sRfermion} in the defining expression for $\smash{\hat{C}}$ [Eq.~\eqref{sdefineC}], which gives $\smash{\hat{R} \hat{C} \hat{R} = \hat{C}}$. To find how the eigenstates transform under $\smash{\hat{R}}$, we use Eqs.~\eqref{sRfermion} and \eqref{seigenmodes} to obtain
\begin{equation}
\hat{R} \hat{f}_0^{\dagger} \hat{R} = \hat{f}_0^{\dagger} \hat{P} \quad \text{and} \quad \hat{R}\push \hat{a}_{k,\pm}^{\dagger} \hat{R} = \pm \hat{a}_{k,\pm}^{\dagger} \hat{P}\;,
\label{sReigenmodes}
\end{equation}
where $\smash{\hat{P}}$ is the total particle-number parity. Using the above relations in Eq.~\eqref{seigenC} and employing the anticommutation between $\smash{\hat{P}}$ and the fermion operators, one finds the state $\smash{|\{\nu_{k,\pm}\},n_0\rangle}$ is an eigenstate of $\smash{\hat{R}}$ with eigenvalue
\begin{equation}
r = (-1)^{N_- + N(N-1)/2}\push,
\label{srpartityint}
\end{equation}
where $N = n_0+\sum_{k} (\nu_{k,+} + \nu_{k,-})$ is the total particle number, and $N_- := \sum_k \nu_{k,-}$ is the number of negatively charged Bell pairs. This expression can be simplified further to yield
\begin{equation}
r = (-1)^{\left\lfloor\frac{1}{2}\left(|\lambda| + \frac{1}{2}\right)\right\rfloor}\;,
\label{srparity}
\end{equation}
where $\lfloor x\rfloor$ is the floor function. Hence, the parity is set by the eigenvalues of $\hat{C}$. As described in the main text, there are $2(l+1)$ distinct eigenvalues, $\lambda\in\{\pm (\eta + 1/2):\eta=0,\dots,l\}$. From Eq.~\eqref{srparity} it follows the reflection parity is even ($r=+1$) if $\eta$ is of the form $4m$ or $4m+3$ for integer $m$, and odd otherwise.

\section{\label{steadystateprop}Properties of steady states}
In this section we derive analytic expressions for some key properties of the steady states, including single-particle and density-density correlations which can be directly measured in experiments. We recall from the main article that the dynamics are decoupled into the $l+1$ distinct eigenspaces of $\smash{\hat{C}^2}$, labeled by $\eta = 0,\dots,l,$ each having a unique steady state as long as both pump and loss rates are nonzero, $\gamma_{\pm} \neq 0$. The steady states are given by
\begin{equation}
\hat{\rho}_{\eta} = \frac{(\gamma_+/\gamma_-\pull )^{\hat{N}} \hat{P}_{\eta}}{\text{Tr}\big[(\gamma_+/\gamma_-\pull )^{\hat{N}} \hat{P}_{\eta}\big]}\;,
\label{ssteadystates}
\end{equation}
where $\smash{\hat{P}_{\eta}}$ is the projector onto the corresponding eigenspace, and $\smash{\hat{N}}$ measures the total particle number. As $\smash{\hat{C}^2}$ does not act on the center site, $\hat{P}_{\eta}$ has the form $\smash{\hat{P}_{\eta} = \hat{Q}_{\eta} \otimes (|0\rangle\langle 0| + |1\rangle\langle 1|)}$, where $|0\rangle$ and $|1\rangle$ describe the center site and $\smash{\hat{Q}_{\eta}}$ encodes the other sites. Within the respective eigenspace, $\hat{\rho}_{\eta}$ is analogous to an infinite-temperature state with a chemical potential, $\smash{\hat{\rho}_{\eta} = e^{\mu\hat{N}} \hat{P}_{\eta} / \mathcal{Z}}$, where $\mu := \ln(\gamma_+/\gamma_-)$, and $\smash{\mathcal{Z} := \text{Tr}\big[(\gamma_+/\gamma_-\pull )^{\hat{N}} \hat{P}_{\eta}\big]}$ is the grand-canonical partition function. Below we calculate this function, followed by closed-form expressions for the densities and two-site correlations. We also compute the relative entropy of coherence \cite{sBaumgratz2014}, a measure that has been put forward to quantify useful coherence in a quantum state, complementary to measures of entanglement.

\subsection{\label{partitionfn}Partition function}
Here we calculate the partition function $\smash{\mathcal{Z} := \text{Tr}[(\gamma_+/\gamma_-\pull )^{\hat{N}} \hat{P}_{\eta}]}$, which will be useful in obtaining the single-particle correlations. First, recall that $\smash{\hat{P}_{\eta}}$ projects onto the eigenspace of $\smash{\hat{C}^2}$ with eigenvalue $(\eta+1/2)^2$, spanned by eigenstates $\smash{|\{\nu_{k,\pm}\},n_0\rangle}$ in Eq.~\eqref{seigenC} which satisfy $\smash{\nu+n_0-1/2 = \pm (\eta +1/2)}$, where $\smash{\nu:=\sum_{k} \nu_{k,+}\pull - \nu_{k,-}\pull}$. The partition function counts these states weighted by a factor $\smash{(\gamma_+/\gamma_-\pull )^N}$, where $N$ is the total particle number, $N = n_0 + \sum_k \nu_{k,+}\pull + \nu_{k,-}$. Since $\smash{\hat{P}_{\eta}}$ is of the form $\smash{\hat{P}_{\eta} = \hat{Q}_{\eta} \otimes (|0\rangle\langle 0| + |1\rangle\langle 1|)}$, it suffices to count only those states where the center site is empty. We call this count $\smash{\mathcal{Z}_0}$. Then the full partition function is given by $\smash{\mathcal{Z} = (1+\gamma_+/\gamma_-)\mathcal{Z}_0}$. To find $\smash{\mathcal{Z}_0}$, we add up states that have $\nu = \eta+1$ or $\nu = -\eta$. All of these states are composed of Bell pairs with occupation $\nu_{k,\pm}\in\{0,1\}$ at positions $k=1,\dots,l$. To count all possibilities, we represent the Bell pairs by a polynomial $\smash{x^{\nu_{k,+} - \nu_{k,-}} \push y^{\nu_{k,+} + \nu_{k,-}}}$ with $y :=\gamma_+/\gamma_-$, such that the powers of $x$ and $y$ measure the contribution toward $\nu$ and $N$, respectively. Then the whole chain is described by the polynomial
\begin{equation}
G(x,y) = \prod_{k=1}^l\; \sum_{\nu_{k,\pm}=0}^1 x^{\nu_{k,+} - \nu_{k,-}} \push y^{\nu_{k,+} + \nu_{k,-}} = \left[1+ y \left(x+\frac{1}{x}\right) + y^2\right]^l ,
\label{sZpoly}
\end{equation}
whose expansion in $x$ plays the role of a generating function for the total charge $\smash{\nu=\sum_{k} \nu_{k,+}\pull - \nu_{k,-}\pull}$. The function $\smash{\mathcal{Z}_0(y)}$ is obtained by adding the coefficients of $x^{\eta+1}$ and $x^{-\eta}$ in this expansion, which can be found in closed form. Rescaling by $1\hspace{-0.03cm} +y$ yields the full partition function
\begin{equation}
\mathcal{Z} =  (1+y) \push y^l \push \mathcal{F}_{l,\eta}(z)\push, \quad \text{with} \quad
\mathcal{F}_{l,\eta}(z):= \left[\push\mathcal{C}_{l+\eta+1}^{(-l)}(z) + \mathcal{C}_{l-\eta}^{(-l)}(z)\right] \;\; \text{and} \;\; 
z := -\frac{1}{2} \left(y+ \frac{1}{y}\right),
\label{sZ}
\end{equation}
where $\smash{\mathcal{C}_n^{(\alpha)}\pull (z)}$ are the Gegenbauer polynomials \cite{sIsmail2005}. For the particle-hole symmetric case, $\gamma_+ = \gamma_-$, $\smash{\mathcal{Z}}$ simply reduces to the degeneracy of the eigenspace, $\smash{\mathcal{Z}(y\to 1) = \text{Tr} (\hat{P}_{\eta}) =  2\binom{2l+1}{l-\eta}}$. For $0 < y \ll 1$, $\smash{\mathcal{Z} \approx \binom{l}{\eta}\push y^{\eta}}$.

\subsection{\label{correlation}Single-particle density matrix}
Here we find closed-form expressions for the single-particle density matrix $\smash{\langle \hat{b}_i^{\dagger} \hat{b}_j\rangle}$ for the steady states $\hat{\rho}_{\eta}$ in Eq.~\eqref{ssteadystates}. As we argued in Sec.~\ref{stringorder}, only the diagonal and antidiagonal elements are allowed to be nonzero. Below we calculate these elements in different symmetry sectors $\eta$ as a function of the pump-to-loss ratio $y:=\gamma_+/\gamma_-$.

\subsubsection{\label{siteocc}Site occupations}
We first focus on the occupations $\smash{n_i := \langle \hat{b}_i^{\dagger} \hat{b}_i\rangle}$. Note that $\hat{\rho}_{\eta}$ has a product form $\smash{\hat{\rho}_{\eta} = \hat{\rho}_{\eta}^{\prime} \otimes \hat{\rho}^{(0)}}$, where $\smash{\hat{\rho}^{(0)}}$ describes the center site, $\smash{\hat{\rho}^{(0)} = (\gamma_+ |1\rangle\langle 1| + \gamma_- |0\rangle\langle 0|)/(\gamma_+\pull + \gamma_-)}$, and $\smash{\hat{\rho}_{\eta}^{\prime}}$ is the reduced density matrix for the other sites. Thus, we conclude the center has occupation
\begin{equation}
n_0 = \gamma_+/(\gamma_+ \pull + \gamma_-) = y/(y+1)\;.
\label{sn0}
\end{equation}
To find $\smash{n_{i\neq 0}}$, recall that $\smash{\hat{P}_{\eta}}$ projects onto the eigenstates $\smash{|\{\nu_{k,\pm}\},n_0\rangle}$ with occupations $n_k = (\nu_{k,+} \pull + \nu_{k,-})/2$ [Eq.~\eqref{snkeigen}], where the pair occupations $\smash{\nu_{k,\pm}}$ at different sites contribute equally toward the eigenvalue. Thus, in the steady state $\hat{\rho}_{\eta}$, all site occupations $\smash{n_{i\neq 0}}$ are identical and can be related to the total particle number as $\smash{n_{i\neq 0} = (\langle\hat{N}\rangle - n_0)/(2l)}$. One can obtain $\smash{\langle\hat{N}\rangle}$ from the partition function $\smash{\mathcal{Z} := \text{Tr}[(\gamma_+/\gamma_-\pull )^{\hat{N}} \hat{P}_{\eta}]}$,
\begin{equation}
\langle\hat{N}\rangle = \frac{\text{Tr}\big[\hat{N} y^{\hat{N}} \hat{P}_{\eta}\big]}{\text{Tr}\big[y^{\hat{N}} \hat{P}_{\eta}\big]} = \frac{y}{\mathcal{Z}} \frac{d\mathcal{Z}}{dy}\;.
\label{stotalocc}
\end{equation}
Using the expression for $\smash{\mathcal{Z}}$ in Eq.~\eqref{sZ} and properties of Gegenbauer polynomials \cite{sIsmail2005}, we find
\begin{equation}
n_{i\neq 0} = \frac{1}{2} + \frac{1}{2} \left(y - \frac{1}{y}\right) \frac{\mathcal{F}_{l-1,\eta}(z)}{\mathcal{F}_{l,\eta}(z)}\push,
\label{snk}
\end{equation}
where $y:= \gamma_+/\gamma_-$, as before. Note the sites are half filled for equal pump and loss, $y=1$, and their occupations grow with the pump-to-loss ratio, as expected. Further, exchanging the pump and loss rates, $y \leftrightarrow 1/y$, exchanges the particle and hole occupations, $n_i \leftrightarrow 1 - n_i$. As shown in Fig.~\ref{occupationfig}(a), $n_{i\neq 0}$ grows monotonically from $\eta/(2l)$ for $y \to 0$ to $1-\eta/(2l)$ for $y\to \infty$. The maximally entangled sector, $\eta = l$, is always half filled as it contains a single particle-hole pair at all reflection-symmetric sites $k$ and $-k$.

\subsubsection{\label{endtoendcorr}End-to-end correlation}

\begin{figure}
\includegraphics[width=0.99\columnwidth]{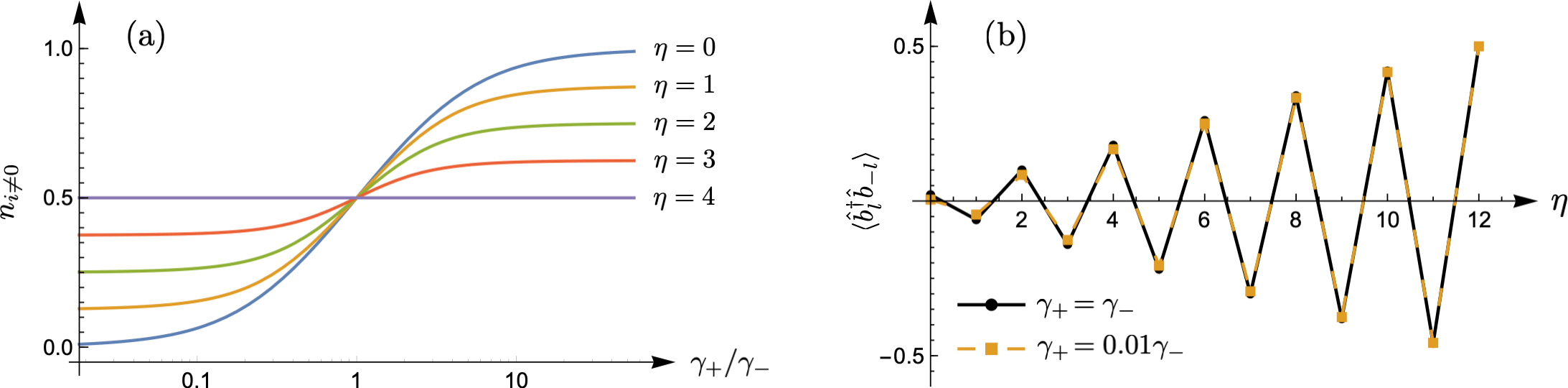}
\caption{\label{occupationfig}(a) Steady-state occupation at sites $i\neq 0$ in different symmetry sectors labeled by $\eta$ as a function of the pump-to-loss ratio $\gamma_+/\gamma_-$ for $l=4$. (b) End-to-end correlation in different sectors for widely varying pump-to-loss ratio for $l=12$, reproduced from Fig.~1(c) in the main article.}
\end{figure}

Now we consider the end-to-end correlation $\smash{\langle\hat{b}_l^{\dagger} \hat{b}_{-l}\rangle}$ which was discussed in the main text as an observable that can distinguish the different steady states for all (nonzero) pump and loss rates. For a given eigenstate $\smash{|\{\nu_{k,\pm}\},n_0\rangle}$, it has the expression $\smash{\langle\hat{b}_l^{\dagger} \hat{b}_{-l}\rangle = (-1)^{\sum_{|i|<l} n_i} (\nu_{l,+}\pull - \nu_{l,-})/2}$ [see Eq.~\eqref{sstringorder}]. To find the correlation in the steady state $\hat{\rho}_{\eta}$, one has to sum over all eigenstates that fall into either of four categories, which give the same eigenvalue $(\eta+1/2)^2$, (i) $n_0 = 0, \nu=\eta+1$, (ii) $n_0=0, \nu=-\eta$, (iii) $n_0=1, \nu = \eta$, and (iv) $n_0=1, \nu = -(\eta+1)$, where $\smash{\nu:=\sum_{k} \nu_{k,+}\pull - \nu_{k,-}\pull}$. However, those in groups (ii) and (iii) [or (i) and (iv)] are related one-to-one by swapping the occupations $\smash{\nu_{k,+} \pull\leftrightarrow \nu_{k,-}}$. Hence, their contributions differ only by a factor of $y$ due to one extra particle at the center for the latter group, so we can write $\smash{\langle\hat{b}_l^{\dagger} \hat{b}_{-l}\rangle = (1+y) \langle\hat{b}_l^{\dagger} \hat{b}_{-l}\rangle_0}$, where the subscript 0 denotes the sum over groups (i) and (ii) only. To evaluate this sum, we follow the procedure in Sec.~\ref{partitionfn} and represent the Bell pairs by a polynomial $\smash{x^{\nu_{k,+} - \nu_{k,-}} \push (-y)^{\nu_{k,+} + \nu_{k,-}}}$ for $1 \leq k < l$, and by $\smash{[(\nu_{l,+}\pull - \nu_{l,-})/2]\push x^{\nu_{l,+} - \nu_{l,-}} \push y^{\nu_{l,+} + \nu_{l,-}}}$ for $k=l$. These terms are designed so that the powers of $x$ and $y$ keep track of the partial sums toward $\nu$ and $N$, respectively, and the other factors measure the correlation. Then summing over all occupations $\smash{\{\nu_{k,\pm}\}}$ yields the polynomial for the whole chain,
\begin{equation}
G_l (x,y) = \frac{y}{2}\left( x - \frac{1}{x} \right) \left[1- y\left(x+\frac{1}{x}\right) + y^2\right]^{l-1}.
\label{sGl}
\end{equation}
We find $\smash{\langle\hat{b}_l^{\dagger} \hat{b}_{-l}\rangle_0}$ by adding the coefficients of $x^{\eta+1}$ and $x^{-\eta}$ in the expansion of $G_l(x,y)$ and dividing by the partition function $\mathcal{Z}$ in Eq.~\eqref{sZ}. Multiplying the result by $1+y$ gives the full correlation
\begin{equation}
\langle\hat{b}_l^{\dagger} \hat{b}_{-l}\rangle = \frac{(-1)^{\eta}}{2} \left[\frac{\mathcal{F}_{l-1,\eta-1}(z) - \mathcal{F}_{l-1,\eta+1}(z)}{\mathcal{F}_{l,\eta}(z)}\right],
\label{sendtoend}
\end{equation}
with $z:= -1/2\push (y+1/y)$, as before. Note the correlation is unaffected by exchanging pump and loss rates, $y \leftrightarrow 1/y$. For equal pump and loss, $y=1$, it reduces to the simple expression
\begin{equation}
\langle\hat{b}_l^{\dagger} \hat{b}_{-l}\rangle \big\vert_{\gamma_+ = \gamma_-} = \frac{(-1)^{\eta}}{2} \left(\frac{2\eta+1}{2l+1}\right)\push,
\label{sendtoendequal}
\end{equation}
which shows the correlation magnitude grows uniformly with the sector label $\eta$, with $\smash{\langle\hat{b}_l^{\dagger} \hat{b}_{-l}\rangle = (-1)^{l}/2}$ for $\eta = l$, the maximally entangled sector. Similarly, in the limit $y\to 0$ or $y\to \infty$, we find
\begin{equation}
\langle\hat{b}_l^{\dagger} \hat{b}_{-l}\rangle \big\vert_{\gamma_{\pm}/\gamma_{\mp} \to 0} = \frac{(-1)^{\eta}}{2}\push \frac{\eta}{l}\;.
\label{sendtoend0}
\end{equation}
As shown in Fig.~1(c) of the main text, reproduced in Fig.~\ref{occupationfig}(b), the correlation is relatively insensitive to $y$ throughout these regimes, but can perfectly distinguish the steady states $\hat{\rho}_{\eta}$ from one another.

\subsubsection{\label{othercorr}Other antidiagonal correlations}
\begin{figure}
\includegraphics[width=0.99\columnwidth]{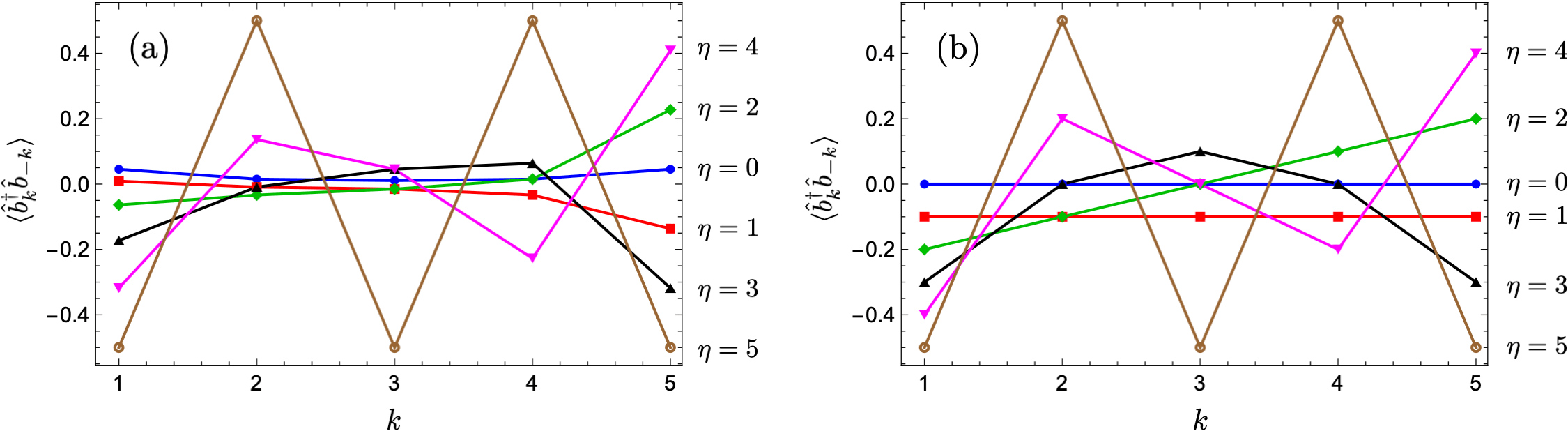}
\caption{\label{othercorrfig}Correlation between sites $k$ and $-k$ in different symmetry sectors $\eta$ for $l=5$, with (a) $\gamma_+/\gamma_- = 1$ and (b) $\gamma_+/\gamma_- \to 0$.}
\end{figure}

In general, correlations of the form $\smash{\langle\hat{b}_k^{\dagger} \hat{b}_{-k}\rangle}$ can be obtained by following the same line of reasoning as for $k=l$ in Sec.~\ref{endtoendcorr}. For an eigenstate $\smash{|\{\nu_{k,\pm}\},n_0\rangle}$, these are given by $\smash{\langle\hat{b}_k^{\dagger} \hat{b}_{-k}\rangle = (-1)^{\sum_{|i|<k} n_i} (\nu_{k,+}\pull - \nu_{k,-})/2}$ [see Eq.~\eqref{sstringorder}]. To sum over all states in a given eigenvalue sector, we represent the Bell pairs by polynomials $\smash{x^{\nu_{i,+} - \nu_{i,-}} \push (-y)^{\nu_{i,+} + \nu_{i,-}}}$ for $1\leq i <k$, $\smash{[(\nu_{k,+}\pull - \nu_{k,-})/2]\push x^{\nu_{k,+} - \nu_{k,-}} \push y^{\nu_{k,+} + \nu_{k,-}}}$ for $i=k$, and $\smash{x^{\nu_{i,+} - \nu_{i,-}} \push y^{\nu_{i,+} + \nu_{i,-}}}$ for $i>k$. As in Sec.~\ref{endtoendcorr}, the polynomials are designed to measure the contribution toward $\smash{\langle\hat{b}_k^{\dagger} \hat{b}_{-k}\rangle}$ while keeping track of the eigenvalue and particle number. Summing over all occupations $\smash{\{\nu_{i,\pm}\}}$ gives the polynomial for the whole chain,
\begin{equation}
G_k (x,y) = \frac{y}{2}\left( x - \frac{1}{x} \right) \left[1- y\left(x+\frac{1}{x}\right) + y^2\right]^{k-1} \left[1+ y\left(x+\frac{1}{x}\right) + y^2\right]^{l-k}.
\label{sGk}
\end{equation}
We define $\smash{T^{l,k}_{\nu}(y)}$ as the coefficient of $x^{\nu}$ in the expansion of $\smash{G_k(x,y)}$. Then the steady-state correlation is given by
\begin{equation}
\langle\hat{b}_k^{\dagger} \hat{b}_{-k}\rangle = \frac{1+y}{\mathcal{Z}}\push \big[T^{l,k}_{\eta+1}(y) + T^{l,k}_{-\eta}(y)\big] = y^{-l} \left[\frac{T^{l,k}_{\eta+1}(y) + T^{l,k}_{-\eta}(y)}{\mathcal{F}_{l,\eta}(z)}\right]\push,
\label{santidiag}
\end{equation}
where we have substituted $\smash{\mathcal{Z}}$ from Eq.~\eqref{sZ} with $z:= -1/2\push (y+1/y)$. Incorporating the $y^{-l}$ factor into $\smash{G_k(x,y)}$, it can be shown that $\smash{\langle\hat{b}_k^{\dagger} \hat{b}_{-k}\rangle}$ is purely a function of $z$, so exchanging pump and loss rates does not affect the correlation, as we found in the last section. Equation~\eqref{santidiag} simplifies for equal pump and loss rates, or $y=1$. Then $\smash{\mathcal{Z} =  2\binom{L}{l-\eta}}$ and the coefficients $\smash{T^{l,k}_{\nu}(1)}$ have a closed form, $\smash{T^{l,k}_{\nu}(1) =\big(\bar{T}^{k-1,l-k}_{\nu-1} \pull - \bar{T}^{k-1,l-k}_{\nu+1}\big)/2}$, where
\vspace{0.1cm}
\begin{equation}
\bar{T}^{p,q}_{\nu} :=
\begin{cases}
(-1)^p \binom{2q}{p+q+\nu} \; {}_2F_1\pull\left(-2p,\push -\nu-p-q;\push 1+q-p-\nu;\push  -1 \right), & \nu \leq q-p \\[8pt]
(-1)^{q+\nu} \binom{2p}{p+q-\nu}\; {}_2F_1\pull\left(-2q,\push \nu -p-q;\push 1+p-q+\nu;\push  -1 \right), & \nu > q-p \;,
\end{cases}
\label{scoefequal}
\vspace{0.1cm}
\end{equation}
with ${}_2F_1$ denoting the ordinary hypergeometric function. Either expression in Eq.~\eqref{scoefequal} works for all $\nu$ provided one evaluates $\smash{\lim_{\nu^{\prime}\to\nu}\bar{T}^{p,q}_{\nu^{\prime}}}$. Similarly, for $y\to 0$ or $y\to\infty$, we find
\begin{equation}
\langle\hat{b}_k^{\dagger} \hat{b}_{-k}\rangle \big\vert_{\gamma_{\pm}/\gamma_{\mp} \to 0} =
\begin{cases}
\frac{(-1)^k}{2} \; {}_2F_1\pull\left(\eta-l, 1-k;\push \eta-k+1;\push -1 \right) \binom{l-k}{\eta - k}  / \binom{l}{\eta}, & k \leq \eta \\[8pt]
\frac{(-1)^{\eta}}{2} \; {}_2F_1\pull\left(k-l, 1-\eta;\push k-\eta+1;\push -1 \right) \binom{k-1}{\eta-1}  / \binom{l}{\eta}, & k > \eta\;.
\end{cases}
\end{equation}
Figure~\ref{othercorrfig} shows how the correlations vary in both regimes. Note the maximally entangled sector, with $\eta=l$, always oscillates between $\pm 1/2$, and the end-to-end correlation grows steadily with $\eta$, as found in Sec.~\ref{endtoendcorr}.

\subsection{\label{nncorr}Density-density correlations}

\begin{figure}
\includegraphics[width=0.85\columnwidth]{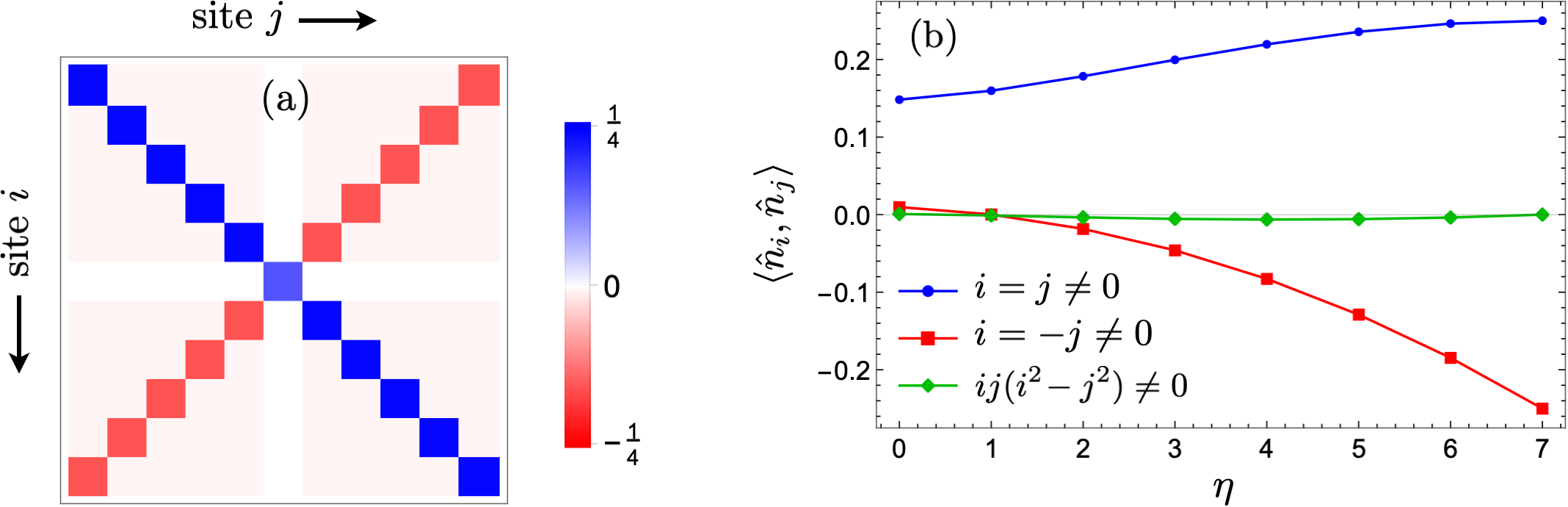}
\caption{\label{nncorrfig}(a) Density-density correlations $\langle \hat{n}_i, \hat{n}_j \rangle := \langle \hat{n}_i \hat{n}_j \rangle - n_i n_j$ in the steady state $\hat{\rho}_{\eta}$ for $l=5$, $\eta=4$, and $\gamma_+/\gamma_-=0.25$. (b) Nonzero correlations in different symmetry sectors $\eta$ for $l=7$ and $\gamma_+/\gamma_-=0.25$.}
\end{figure}

Signatures of the steady states also appear in the density-density correlations $\langle \hat{n}_i, \hat{n}_j \rangle := \langle \hat{n}_i \hat{n}_j \rangle - n_i n_j$, which can be found in closed form following the procedure in Sec.~\ref{correlation}. As shown in Fig.~\ref{nncorrfig}, these correlations are dominated by the terms $i=\pm j$, similar to the one-particle correlations. In particular, the center site is uncorrelated with all the other sites, $\langle \hat{n}_0, \hat{n}_k \rangle_{k \neq 0} = 0$, as expected, with $\langle \hat{n}_0, \hat{n}_0 \rangle_{k \neq 0} = \gamma_+ \gamma_- / (\gamma_+ \pull + \gamma_-)^2$. Additionally, one finds
\begin{align}
\langle \hat{n}_k, \hat{n}_k \rangle_{k\neq 0} &= \frac{1}{4} - (z^2-1) \left[ \frac{\mathcal{F}_{l-1,\eta}(z)}{\mathcal{F}_{l,\eta}(z)} \right]^2 ,
\label{snndiag}\\
\langle \hat{n}_k, \hat{n}_{-k} \rangle_{k\neq 0} &= - \frac{1}{4} - z\push \frac{\mathcal{F}_{l-1,\eta}(z)}{\mathcal{F}_{l,\eta}(z)} - (z^2-1) \left[ \frac{\mathcal{F}_{l-1,\eta}(z)}{\mathcal{F}_{l,\eta}(z)} \right]^2 ,
\label{snnantidiag}\\
\text{and} \;\; \langle \hat{n}_i, \hat{n}_j \rangle &= (z^2-1) \left\{ \frac{\mathcal{F}_{l-2,\eta}(z)}{\mathcal{F}_{l,\eta}(z)} - \left[ \frac{\mathcal{F}_{l-1,\eta}(z)}{\mathcal{F}_{l,\eta}(z)} \right]^2 \right\} \;\; \text{for} \;\; i j (i^2 - j^2) \neq 0 \;,
\label{snnij}
\end{align}
where we have used the definitions in Eq.~\eqref{sZ} with $y:= \gamma_+/\gamma_-$. For the maximally entangled sector $\eta=l$, the only (off-center) nonzero elements are $\langle \hat{n}_k, \hat{n}_{-k} \rangle_{k \neq 0} = -\langle \hat{n}_k, \hat{n}_{k} \rangle_{k\neq 0} = -1/4$, exhibiting maximal antidiagonal coherence. Further, the expressions in Eqs.~\eqref{snndiag}--\eqref{snnij} simplify for $\gamma_+ = \gamma_-$, yielding
\begin{equation}
\langle \hat{n}_k, \hat{n}_k \rangle = 1/4\;, \;\; \langle \hat{n}_k, \hat{n}_{-k} \rangle_{k\neq 0} 
= - \frac{1}{4} + \frac{(l-\eta) (l+\eta+1)}{2l (2l+1)} \;,
\end{equation}
and $\langle \hat{n}_i, \hat{n}_j \rangle = 0$ otherwise. Similarly, for $\gamma_+/\gamma_-\to 0$, we obtain
\begin{equation}
\langle \hat{n}_k, \hat{n}_k \rangle_{k\neq 0} = \frac{\eta}{4l} \left(2-\frac{\eta}{l}\right)\push, \;\;
\langle \hat{n}_k, \hat{n}_{-k} \rangle_{k\neq 0} = -\frac{\eta^2}{4l^2} \push,\;\;\text{and}\;\;
\langle \hat{n}_i, \hat{n}_j \rangle = - \frac{\eta(l-\eta)}{4l^2(l-1)} \;\;\text{for}\;\; ij (i^2-j^2)\neq 0\push.
\end{equation}
For all pump-to-loss ratio, the antidiagonal correlation $\langle \hat{n}_k, \hat{n}_{-k} \rangle_{k\neq 0}$ decreases monotonically with $\eta$, reaching $-1/4$ for $\eta=l$ [Fig.~\ref{nncorrfig}(b)]. It can thus be measured to characterize the steady states.

\subsection{\label{entcoherence}Relative entropy of coherence}
A growing number of studies in recent years have been devoted to developing a quantitative theory of coherence as a resource \cite{sStreltsov2017}, following parallel developments in entanglement measures. In particular, a physically well motivated coherence measure for a density operator $\hat{\rho}$ is the relative entropy of coherence \cite{sBaumgratz2014}, defined as
\begin{equation}
S_{\text{rel.ent.}} := S(\hat{\rho}^{\text{diag}}) - S(\hat{\rho})\;,
\label{sdefrelent}
\end{equation}
where $\smash{\hat{\rho}^{\text{diag}}}$ is a diagonal matrix with the diagonal entries of $\smash{\hat{\rho}}$, and $S$ is the von Neumann entropy, $\smash{S(\hat{\rho}):=-\text{Tr}(\hat{\rho} \log \hat{\rho}})$. Clearly, $\smash{C_{\text{rel.ent.}}}$ measures coherence in a preferred basis, which is dictated by the experimental system. For our model of a qubit array, a natural basis is given by the Fock states, $\{0,1\}^{\otimes 2l+1}$, which are the easiest to access experimentally. Here we calculate the relative entropy in this basis for the steady states $\smash{\hat{\rho}_{\eta}}$ in the particle-hole symmetric case, $\gamma_+ \pull = \gamma_-$, finding similar variation as the logarithmic entanglement negativity discussed in the main text.

\begin{figure}
\includegraphics[width=0.92\columnwidth]{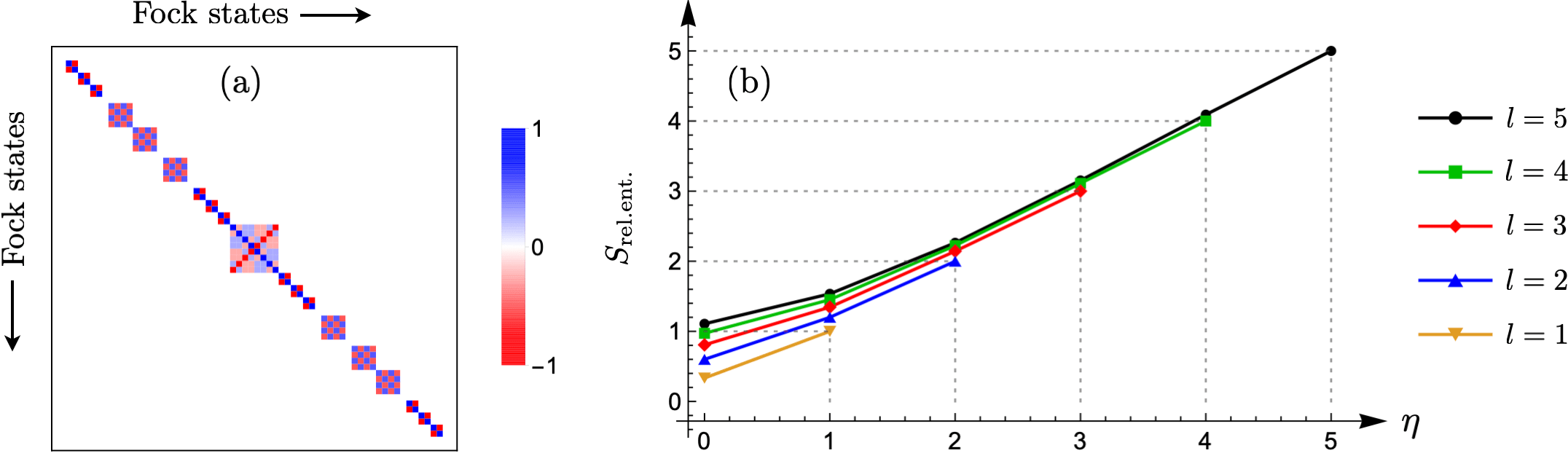}
\caption{\label{relentfig}(a) Rescaled density matrix representing $\smash{\hat{Q}_{\eta}}$ in the basis of Fock states for $l=3$, $\eta=1$. The blocks comprise states with a given number of Bell pairs. (b) Relative entropy of coherence in different symmetry sectors $\eta$ for varying system size $l$.}
\end{figure}

Recall that, for $\gamma_+ = \gamma_-$, $\smash{\hat{\rho}_{\eta}} = \hat{P}_{\eta} / \mathcal{Z}$, where $\smash{\hat{P}_{\eta}}$ is the projector onto the symmetry sector and $\smash{\mathcal{Z} = \text{Tr}(\hat{P}_{\eta}) = 2\binom{2l+1}{l-\eta}}$ [see Eq.~\eqref{ssteadystates}]. Since $\smash{\hat{P}_{\eta}}$ is equivalent to an infinite-temperature state within the sector, the von Neumann entropy is simply given by the dimension $\smash{\mathcal{Z}}$, $\smash{S(\hat{\rho}_{\eta}) = \ln \mathcal{Z}}$. Further, the diagonal elements of $\smash{\hat{\rho}_{\eta}}$ all have the factor $\smash{1/\mathcal{Z}}$, which add up to $\smash{\ln\mathcal{Z}}$ in $\smash{S(\hat{\rho}_{\eta}^{\text{diag}})}$, so we find $\smash{S_{\text{rel.ent}} = S(\hat{P}_{\eta}^{\text{diag}})/\mathcal{Z}}$. To evaluate this entropy, we recall that $\smash{\hat{P}_{\eta}}$ has the form $\smash{\hat{P}_{\eta} = \hat{Q}_{\eta} \otimes (|0\rangle\langle 0| + |1\rangle\langle 1|)}$, where $|0\rangle$ and $|1\rangle$ describe the center site. Thus, $\smash{S(\hat{P}_{\eta}^{\text{diag}}) = 2 S(\hat{Q}_{\eta}^{\text{diag}})}$, and
\begin{equation}
S_{\text{rel.ent.}} = S(\hat{Q}_{\eta}^{\text{diag}})/
\text{\scriptsize $\left(\pull\pull\begin{array}{c}2l+1\\[2pt]l-\eta\end{array}\pull\pull\right)$}\;.
\label{sQrelent}
\end{equation}
The reduced density operator $\smash{\hat{Q}_{\eta}}$ projects onto the eigenstates $\smash{|\{\nu_{k,\pm}\},n_0=0\rangle}$ given in Eq.~\eqref{seigenC}, where $\smash{\nu_{k,\pm}\in\{0,1\}}$ give the occupation of particle-hole pairs with charge $\pm 1$ at sites $k$ and $-k$. These Bell pairs are created by operators $\smash{\hat{a}_{k,\pm}^{\dagger} = (\hat{f}_k^{\dagger} \pm \hat{f}_{-k}^{\dagger})/\sqrt{2}}$ [Eq.~\eqref{seigenmodes}]. It follows that $\smash{\hat{Q}_{\eta}}$ is block diagonal in the Fock states, where each block corresponds to a given distribution of the Bell pairs, as shown in Fig.~\ref{relentfig}(a). To understand this structure, consider an eigenstate with a given total charge $\smash{\nu:=\sum_{k=1}^l \nu_{k,+}\pull - \nu_{k,-}\pull}$. Suppose there are $\nu+2m$ singly-occupied positions $k\in\mathcal{K}$, which contain $m$ negatively charged pairs and $\nu+m$ positively charged pairs. All other positions are either empty or have both positive and negative charges. Such a state contributes a weight $\smash{1/2^{\nu+2m}}$ to all Fock states that have a particle at either $k$ or $-k$ for $k\in\mathcal{K}$, and definite occupations at other sites. There are $\smash{2^{\nu+2m}}$ such Fock states which form a block. Exchanging the locations of a positive charge and a negative charge within $\mathcal{K}$ contributes to the same block. Hence, the total weight of every Fock state within this block is $\smash{\binom{\nu+2m}{m}/2^{\nu+2m}}$, yielding the block entropy
\begin{equation}
S^{\text{diag}}_{\nu,m} = -\binom{\nu+2m}{m} \log\pull\left[\binom{\nu+2m}{m} \frac{1}{2^{\nu+2m}}\right]\push.
\label{sblockrelent}
\end{equation}
There are $\smash{2^{l-\nu-2m}}$ such blocks for a given $\smash{\mathcal{K}}$, corresponding to empty or doubly-occupied states for positions $\smash{k \notin \mathcal{K}}$. Further, choosing a different set $\smash{\mathcal{K}^{\prime}}$ of the same size gives the same entropy, for which there are $\smash{\binom{l}{\nu+2m}}$ possibilities. Thus, we obtain the net relative entropy from eigenstates with a given $\nu \geq 0$,
\begin{equation}
S^{\text{diag}}_{\nu} = -\pull\sum_{m=0}^{(l-\nu)/2} 2^{l-\nu-2m} \binom{l}{\nu+2m} \binom{\nu+2m}{m} \log\pull\left[\binom{\nu+2m}{m} \frac{1}{2^{\nu+2m}}\right]\push.
\label{snurelent}
\end{equation}
For $\nu<0$, the role of positive and negative charges are reversed and one finds the same relative entropy, $\smash{S^{\text{diag}}_{-\nu} \pull=\pull S^{\text{diag}}_{\nu}}$. The operator $\smash{\hat{Q}_{\eta}}$ in Eq.~\eqref{sQrelent} projects onto the span of eigenstates with $\nu=\eta+1$ or $\nu=-\eta$, both of which give the same eigenvalue $(\eta+1/2)^2$ of the symmetry operator $\smash{\hat{C}^2}$. Thus, $\smash{S(\hat{Q}_{\eta}^{\text{diag}}) = S^{\text{diag}}_{\eta+1} + S^{\text{diag}}_{\eta}}$. Substituting in Eq.~\eqref{sQrelent} and rewriting dummy variables, we find the final expression
\begin{equation}
S_{\text{rel.ent.}} = \binom{2l+1}{l-\eta}^{\pull\pull -1} \sum_{m=\eta}^{l} 2^{l-m} \binom{l}{m} \binom{m}{\left\lfloor \frac{m-\eta}{2}\right\rfloor} \log\pull \left[2^m \bigg/\binom{m}{\left\lfloor \frac{m-\eta}{2}\right\rfloor}\right]\push,
\label{srelent}
\end{equation}
where $\smash{\lfloor x \rfloor}$ is the floor function, as before. For the maximally entangled sector, $\eta=l$, the above expression reduces to $\smash{S_{\text{rel.ent.}} = l \log2}$. As shown in Fig.~\ref{relentfig}(b), with log base 2, the relative entropy grows monotonically with $\eta$, similar to the logarithmic negativity plotted in Fig.~2 of the main text.

\section{\label{protocol}Preparation protocol}
As described in the main article, the steady states in different symmetry sectors can be selectively prepared if one can engineer loss of the JW fermions from the center site. First, one uses only the JW fermion loss to drive the system from a symmetric Fock state to a pure state with Bell pairs in a given sector $\eta$. Second, one switches from the fermion loss to the boson pump and loss, driving the system to the steady state $\smash{\hat{\rho}_{\eta}}$ in Eq.~\eqref{ssteadystates}. In this section we derive an expression of the symmetry operator $\smash{\hat{C}}$ in terms of the occupations of even and odd single-particle modes, leading to the mapping between symmetric Fock states and the sector index $\eta$. We also analyze the timescales for preparation, extract optimal parameters, and simulate an experimental setting with dissipation on all sites.

\subsection{\label{spmodes}Even and odd single-particle modes}
The single-particle eigenmodes, $\hat{F}_{m}$, of the Hamiltonian in Eq.~\eqref{sfermionhamil} can be found by requiring $\smash{[\hat{H},\hat{F}_{m}^{\dagger}]=\varepsilon_{m} \hat{F}_{m}^{\dagger}}$, which gives $l$ odd modes and $l+1$ even modes of the JW fermions. They are given by
\begin{equation}
\hat{F}_{m} = \sqrt{\frac{1}{l+1}} \;\sum_{j=-l}^l \sin\pull \left[\frac{\pi m j}{2(l+1)}+\frac{\pi m}{2}\right] \hat{f}_{j}\;, \quad \text{and} \quad \varepsilon_{m} = -2 \hbar J \cos\left[\frac{\pi m}{2(l+1)}\right],
\label{sspmodes}
\end{equation}
where $m\in\{2,4,\dots,2l\}$ for the odd modes and $m\in\{1,3,\dots,2l+1\}$ for the even modes. Using these expressions, one can find the total occupation of the even and odd modes,
\begin{equation}
\hat{N}_{\text{even}} = \hat{n}_0 + \sum_{k=1}^l \hat{a}_{k,+}^{\dagger} \hat{a}_{k,+} \;, \quad\text{and} \quad \hat{N}_{\text{odd}} = \sum_{k=1}^l \hat{a}_{k,-}^{\dagger} \hat{a}_{k,-} \;,
\label{sNevenodd}
\end{equation}
where the operators $\hat{a}_{k,\pm}$ are defined in Eq.~\eqref{seigenmodes}. Comparing with Eq.~\eqref{sCdiag}, we find $\hat{C} = \hat{N}_{\text{even}} - \hat{N}_{\text{odd}}-1/2$. Thus, the symmetry sectors are characterized by a definite value of $\smash{(\hat{N}_{\text{even}} \pull-\pull \hat{N}_{\text{odd}}\pull-\pull 1/2)^2}$. In particular, states that have a given number of particles in the odd modes and none in the even modes belong to the sector $\smash{\eta=N_{\text{odd}}}$.

\subsection{\label{fockmapping}Mapping Fock states to symmetry sectors}
As explained in the main text, the first stage of the protocol uses JW fermion loss at the center, for which all the odd modes are unaffected and all the even modes die out. Thus, any initial state with a definite odd-mode occupation will be driven toward the sector with $\smash{\eta = N_{\text{odd}}}$. Below we show this is exemplified by symmetric Fock states.

Consider a symmetric Fock state of the bosons,
\begin{equation}
|\{n_k\}\rangle := \big(\hat{b}_0^{\dagger}\big)^{\pull n_0}\prod_{k=1}^l \big(\hat{b}_k^{\dagger} \hat{b}_{-k}^{\dagger}\big)^{\pull n_k} \push |0\rangle \;,
\label{sfocksym}
\end{equation}
where $n_k \in \{0,1\}$. Using the transformation in Eq.~\eqref{sJordanWigner}, one finds such a state is also a Fock state of the JW fermions with the same occupations, $|\{n_k\}\rangle =\pm \big(\hat{f}_0^{\dagger}\big)^{\pull n_0}\pull\prod_{k=1}^l \pull\big(\hat{f}_k^{\dagger} \hat{f}_{-k}^{\dagger}\big)^{\pull n_k} |0\rangle$. Next, using $\smash{\hat{f}_k^{\dagger} \hat{f}_{-k}^{\dagger} = \hat{a}_{k,-}^{\dagger} \hat{a}_{k,+}^{\dagger}}$ from Eq.~\eqref{seigenmodes} and the results in Eq.~\eqref{sNevenodd} yields $N_{\text{odd}} = \sum_{k=1}^l n_k$. Thus, a symmetric Fock state of the form in Eq.~\eqref{sfocksym} will be driven toward the sector $\eta = \sum_{k=1}^l n_k$, set by the initial occupations.

\subsection{\label{timescales}Timescales and optimal parameters}
In experiments, the observation timescales are limited by the presence of residual dissipation, on-site disorder, or other unwanted energy scales. Here we estimate the time required for implementing both stages of our preparation protocol, and extract optimal parameters which give the fastest preparation time.

We first briefly review how the dynamics converge in the presence of a general (Markovian) dissipation. As described in the main text, the dynamics are governed by a master equation for the density operator $\hat{\rho}$,
\begin{equation}
\frac{d\hat{\rho}}{dt} = \mathcal{L}\hat{\rho} := -\frac{i}{\hbar}\push [\hat{H},\hat{\rho}] \push + \sum_{\alpha}\pull \hat{L}_{\alpha} \hat{\rho}\hat{L}_{\alpha}^{\dagger} - \frac{1}{2} \{\hat{L}_{\alpha}^{\dagger} \hat{L}_{\alpha}, \hat{\rho}\}\;,
\label{smastereqn}
\end{equation}
where the jump operators $\smash{\hat{L}_{\alpha}}$ model the dissipation, and the Liouvillian $\mathcal{L}$ defines a completely positive trace-preserving map on the set of density operators. If $\smash{D}$ is the dimension of the Hilbert space, $\smash{\mathcal{L}}$ can be represented by a $\smash{D^2 \times D^2}$ matrix that acts on the $\smash{D^2}$ elements of $\hat{\rho}$. Then the solution to Eq.~\eqref{smastereqn} is given by $\smash{|\rho(t)\rangle\rangle} = \sum_r e^{\Lambda_r t} |v_r\rangle\rangle \langle\langle u_r | \rho(0)\rangle\rangle$, where $\smash{\Lambda_r}$ are the eigenvalues of $\smash{\mathcal{L}}$, $\smash{\langle\langle u_r|}$ and $\smash{|v_r\rangle\rangle}$ are the corresponding left and right eigenvectors, and $|\rho(t)\rangle\rangle$ is the vector obtained by flattening the density matrix. The eigenvalues have nonpositive real parts which give decay rates of the associated eigenvectors \cite{sAlbert2014}. Accordingly, the dynamics converge on a timescale set by the eigenvalue with the smallest nonzero decay rate, called the spectral gap, $\smash{\Delta := \text{min}_r \push |\text{Re } \Lambda_r|>0}$.

\begin{figure}
\includegraphics[width=0.99\columnwidth]{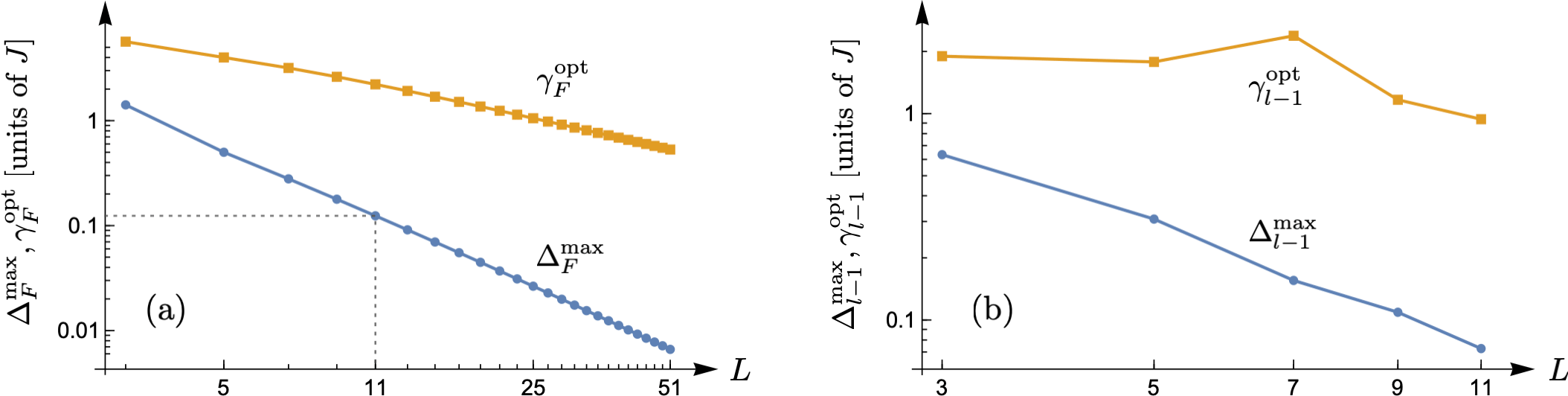}
\caption{\label{timescalefig}(a) Maximum spectral gap $\smash{\Delta_F^{\text{max}}}$ and the corresponding loss rate of JW fermions, $\smash{\gamma_F^{\text{opt}}}$, in the first stage of the protocol as a function of the number of sites, $L=2l+1$, using the free-fermion method of Ref.~\cite{sProsen2008}. (b) Maximum spectral gap $\smash{\Delta_{\eta}^{\text{max}}}$ and optimal pump/loss rate of bosons, $\smash{\gamma_{\eta}^{\text{opt}}}$, in the second stage of the protocol for $\eta = l-1$ and $\gamma_+=\gamma_-:=\gamma$, using exact diagonalization.}
\end{figure}

As explained in Sec.~\ref{fockmapping}, the first stage of our protocol uses JW fermion loss at the center site to drive the system to a given symmetry sector. Such a process is modeled by a single jump operator $\smash{\hat{L}_F := \sqrt{\gamma_F} \hat{f}_0}$, where $\smash{\gamma_F}$ is the loss rate. The dissipation does not affect odd fermionic modes, which evolve unitarily with purely imaginary eigenvalues $\Lambda_r$. The desired symmetry sector is reached when all the even modes decay to zero. This decay rate can be estimated from the spectral gap $\smash{\Delta_F}$ of the Liouvillian projected onto the set of even modes, irrespective of the symmetry sector. In fact, as $\smash{\mathcal{L}}$ is quadratic in the JW fermions [see Eqs.~\eqref{sfermionhamil} and \eqref{smastereqn}], $\smash{\Delta_F}$ can be found by diagonalizing a $4L \times 4L$ matrix using the free-fermion method of Ref.~\cite{sProsen2008}, where $L$ is the number of sites, $L=2l+1$. We find $\smash{\Delta_F}$ vanishes for both $\smash{\gamma_F\to 0}$ and $\smash{\gamma_F \to \infty}$, the latter because of the quantum Zeno effect \cite{sPopkov2020}. It reaches a peak $\smash{\Delta_F^{\text{max}}}$ at an intermediate loss rate $\smash{\gamma_F^{\text{opt}}}$. Figure~\ref{timescalefig}(a) shows how the maximum decay rate and the optimal loss rate vary with $L$. Numerically, they fall off at large $L$ as $\smash{\Delta_F^{\text{max}} \sim 1/L^2}$ and $\smash{\gamma_F^{\text{opt}} \sim 1/L}$. However, for $L\leq 11$, as in a recent experiment \cite{sMa2019}, $\smash{\Delta_F^{\text{max}} > 0.1 J}$, or the convergence time $\smash{\tau_F = 1/\Delta_F < 10/J}$. This estimate agrees with the time evolution in Fig.~3 of the main text, and is more than an order-of-magnitude faster than both residual dephasing or on-site disorder in Ref.~\cite{sMa2019}. Thus, by adjusting $\smash{\gamma_F \sim J}$, one can reliably prepare all symmetry sectors.

The second stage of our protocol uses boson pump and loss at the center site to arrive at the steady state $\hat{\rho}_{\eta}$ within a symmetry sector. This process is modeled by two jump operators, $\smash{\hat{L}_{1,2}:=\sqrt{\gamma_{\pm}} \push\hat{b}_0^{(\dagger)}}$, where $\gamma_{\pm}$ are the pump and loss rates, as discussed before. The dynamics are decoupled into the separate symmetry sectors $\eta$, and every sector has a unique steady state $\hat{\rho}_{\eta}$ with $\smash{\Lambda_r = 0}$, with no other purely imaginary eigenvalues. Unlike in the first stage, the spectral gap $\smash{\Delta_{\eta}}$ now depends on the sector. In particular, for the maximally entangled sector, $\eta=l$, we find $\smash{\Delta_l = (\gamma_+\pull + \gamma_-)/2}$. This is because it is spanned by two maximally entangled eigenstates of the Hamiltonian, that are exchanged by the pump and loss. Thus, one can make the convergence arbitrarily fast (or slow) by tuning the pump and loss rates. As $\eta$ is decreased, the steady state $\hat{\rho}_{\eta}$ becomes less entangled, and the sector dimension grows as $\smash{\binom{L}{l-\eta}}$, making $\smash{\Delta_{\eta}}$ less numerically tractable for large $L$. Figure~\ref{timescalefig}(b) shows how the maximum spectral gap $\smash{\Delta_{\eta}^{\text{max}}}$ and the optimal rate $\smash{\gamma_{\eta}^{\text{opt}}}$ vary with $L$ for $\eta = l-1$ in the particle-hole symmetric case, $\smash{\gamma_+ = \gamma_- :=\gamma}$. In general, we find $\smash{\Delta_{\eta}^{\text{max}}}$ falls off as $L$ is increased or $\eta$ is decreased. However, $\smash{\Delta_{\eta}^{\text{max}} \gtrsim 0.1 J}$ for $L \leq 11$, so the dynamics converge in a few tens of tunneling time with $\gamma\sim J$, as in the first stage with JW fermion loss.

\subsection{\label{unwanteddissipation}Effect of dissipation on all sites}

\begin{figure}
\includegraphics[width=0.94\columnwidth]{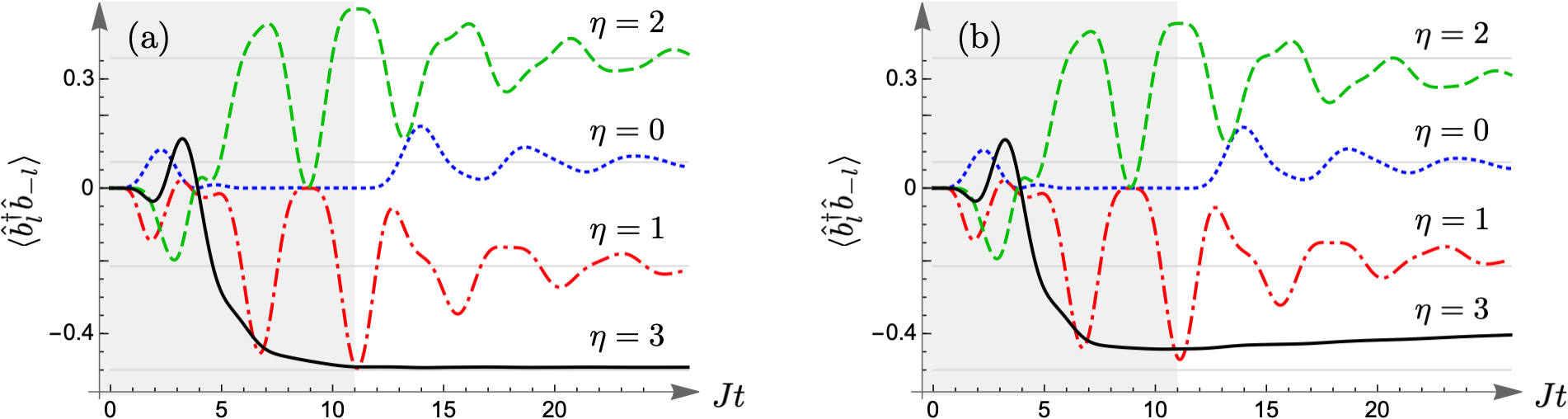}
\caption{\label{fidelityfig}End-to-end coherence during the two-step preparation of the steady states $\hat{\rho}_{\eta}$ for $L=7$, from exact diagonalization. The shaded regions correspond to JW fermion loss at the center with rate $\gamma_{\text{F}} = 3J$, and the white regions correspond to boson pump and loss at the center with rates $\gamma_+ = \gamma_- = 2J$. (a) Idealized case: no dissipation at other sites. (b) Uniform dephasing and loss on all sites, with rates $\gamma_d = J/200$ and $\gamma_l = J/2000$, as in Ref.~\cite{sMa2019}.}
\end{figure}

Here we simulate the preparation protocol in the presence of dephasing and loss on all sites, which are generally present to some degree in experiments. Such dissipation does not preserve the symmetry $\hat{C}^2$, destabilizing the steady states $\hat{\rho}_{\eta}$ at long times. We take the experimental parameters in Ref.~\cite{sMa2019}, where the dephasing and loss rates were $\gamma_d \approx J/200$ and $\gamma_l \approx J/2000$, respectively. In Fig.~\ref{fidelityfig}(b), we plot the evolution of the end-to-end coherence which uniquely characterizes the steady states. Compared to the idealized case with $\gamma_d = \gamma_l = 0$, shown in Fig.~\ref{fidelityfig}(a), we find a fidelity greater than $80\%$ in all of the symmetry sectors, and the states can be clearly distinguished from one another over several tens of tunneling time. Other measures such as entanglement give similar fidelities. This shows our findings can be observed under realistic experimental conditions.

\section{\label{pbc}Extension to periodic boundary}
We assumed open boundary conditions in writing the Hamiltonian in Eq.~\eqref{sbosonhamil}. However, the central results carry over to periodic boundary conditions, which we discuss in this brief section.

\begin{figure}[b]
\includegraphics[width=0.9\columnwidth]{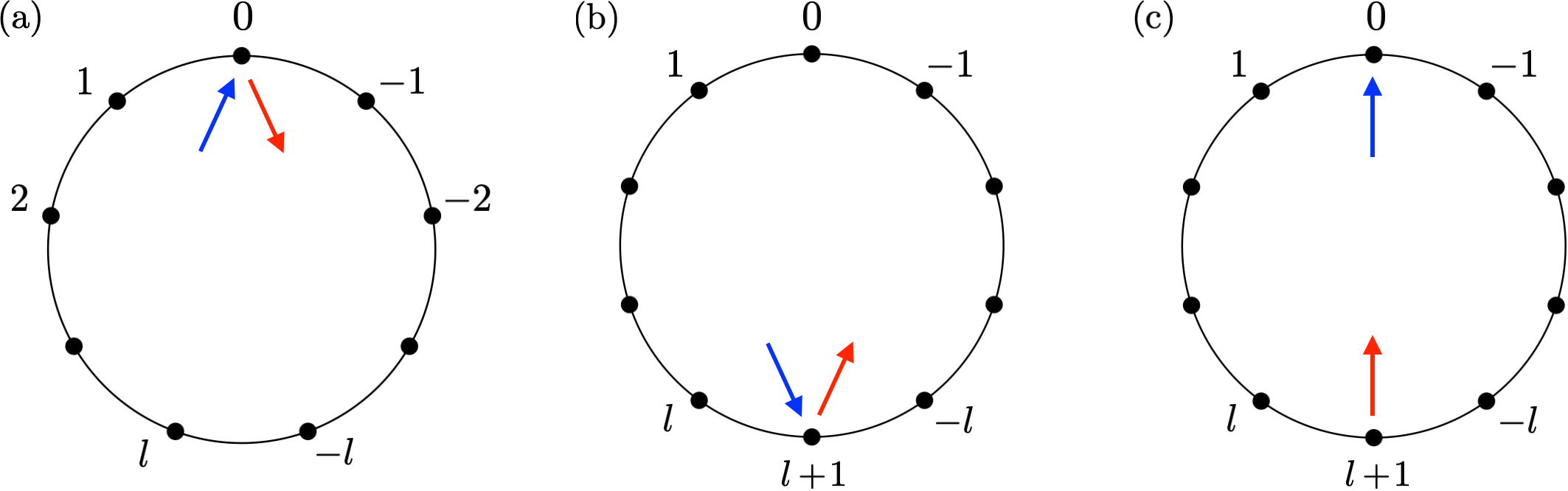}
\caption{\label{pbcfig}Scenarios with periodic boundary where similar results are found: (a) odd number of sites, and (b)--(c) even number of sites with pump and loss occurring at the same site or at diametrically opposite sites.}
\end{figure}

We first consider odd number of sites ($L$), as in the original model. Here the only additional term in the Hamiltonian is $\smash{-\hbar J (\hat{b}_l^{\dagger} \hat{b}_{-l} + \hat{b}_{-l}^{\dagger} \hat{b}_{l})}$, which explicitly commutes with the symmetry operator $\smash{\hat{C}}$ in Eq.~\eqref{sCboson}. Thus, $\smash{\hat{C}^2}$ again generates a strong symmetry for pump and loss at site 0, and one recovers the same steady states. This situation is sketched in Fig.~\ref{pbcfig}(a). Note there is nothing special about site 0; one can construct copies of $\smash{\hat{C}}$ ``centered'' at every site, all of which commute with the Hamiltonian.

When $L$ is even, there is an extra site between $l$ and $-l$, which we label as $l+1$. The Hamiltonian is given by
\begin{equation}
\hat{H}_{\text{even}} = \hat{H} - \hbar J \big(\hat{b}_l^{\dagger} \hat{b}_{l+1} + \hat{b}_{l+1}^{\dagger} \hat{b}_{-l} + \text{H.c.}\big)\;,
\end{equation}
where $\hat{H}$ is defined in Eq.~\eqref{sbosonhamil}. Although $\smash{\hat{H}_{\text{even}}}$ does not commute with $\smash{\hat{C}}$, the symmetry can be generalized as
\begin{equation}
\hat{C}_{\text{even}} = \hat{C} + \frac{1}{2} (-1)^{\sum_{i=-l}^l \hat{n}_i}.
\label{sCeven}
\end{equation}
The new term only depends on the total occupation between sites $-l$ and $l$, and commutes with $\hat{H}$. Therefore,
\begin{equation}
[\hat{H}_{\text{even}}, \hat{C}_{\text{even}}] = \hbar J \big[\hat{C}_{\text{even}}, \hat{b}_l^{\dagger} \hat{b}_{l+1} + \hat{b}_{l+1}^{\dagger} \hat{b}_{-l} \big] -\text{H.c.}\;.
\end{equation}
Substituting for $\hat{C}_{\text{even}}$ from Eqs.~\eqref{sCeven} and \eqref{sCboson}, and using the bosonic commutations, one finds
\begin{equation}
[\hat{H}_{\text{even}}, \hat{C}_{\text{even}}] = \hbar J \push (-1)^{\sum_{|i|<l} \hat{n}_i} \bigg[ \hat{b}_l^{\dagger} \hat{b}_{-l} + \hat{b}_{-l}^{\dagger} \hat{b}_l + \frac{1}{2} (-1)^{\hat{n}_l + \hat{n}_{-l}} ,\; \hat{b}_l^{\dagger} \hat{b}_{l+1} + \hat{b}_{l+1}^{\dagger} \hat{b}_{-l} \bigg]  - \text{H.c.} = 0\;.
\end{equation}
As $\smash{\hat{C}_{\text{even}}}$ does not act on the site $l+1$, it generates a strong symmetry for any local dissipation (e.g., pump and loss) occurring at this site, as shown in Fig.~\ref{pbcfig}(b). Note one can also construct a copy of $\smash{\hat{C}_{\text{even}}}$ ``centered'' at site $l+1$, say $\smash{\hat{C}_{\text{even}}^{\prime}}$, such that $\smash{\hat{C}_{\text{even}}^{\prime 2}}$ gives a strong symmetry. However, these two generators are related by $\smash{\hat{C}_{\text{even}}^{\prime} = -(-1)^{\hat{N}} \hat{C}_{\text{even}}}$, where $\hat{N}$ is the total occupation, thus $\smash{\hat{C}_{\text{even}}^{\prime 2} = \hat{C}_{\text{even}}^2}$. Hence, the dynamics are decoupled into eigenspaces of $\smash{\hat{C}_{\text{even}}}$, leading to multiple steady states. The spectrum of $\smash{\hat{C}_{\text{even}}}$ is composed of eigenstates $\smash{|\{\nu_{k,\pm}\},n_0\rangle}$ and $\smash{\hat{b}_{l+1}^{\dagger} |\{\nu_{k,\pm}\},n_0\rangle}$, as defined in Eq.~\eqref{seigenC}, with eigenvalues
\begin{equation}
\lambda_{\text{even}} = \nu + \big[ (-1)^{\nu+n_0} \pull - (-1)^{n_0}\big] /2\;,
\label{slambdaeven}
\end{equation}
where $\nu \in \{-l,-l+1,\dots,l-1,l\}$ and $n_0 \in \{0,1\}$. As before, the magnitude of $\nu$ gives the number of Bell pairs in the system. From Eq.~\eqref{slambdaeven}, when $\nu$ is even, $\smash{\lambda_{\text{even}} = \nu}$, and when $\nu$ is odd, $\smash{\lambda_{\text{even}} = \nu \pm 1}$ depending on $n_0$. Therefore, the eigenvalues are all even numbers. For odd $l$, or $L=4m$ for integer $m$, there are $l+2$ distinct symmetry sectors, $\smash{\lambda_{\text{even}} \in \{-l-1,-l+1,\dots,l-1,l+1\}}$. Two of these, with $\smash{\lambda_{\text{even}} = \pm (l+1)}$, are maximally entangled. Conversely, for even $l$, or $L=4m+2$, there are $l+1$ sectors, $\smash{\lambda_{\text{even}} \in \{-l,-l+2,\dots,l-2,l\}}$. Here, the maximally entangled states with $\smash{\lambda_{\text{even}} = \pm l}$ are mixed with less entangled ones. For equal pump and loss, the steady states are given by projectors onto the different sectors, as in the original model with open boundaries.

Figure~\ref{pbcfig}(c) depicts another scenario, where the pump and loss occur at diametrically opposite sites. Here, $\smash{\hat{C}_{\text{even}}^2}$ generates a strong symmetry. Consequently, the number of symmetry sectors and steady states are (roughly) halved. However, this does not affect the long-range coherence in the maximally entangled sector for odd $l$.


\begin{thebibliography}{84}
\bibitem [{spi()}]{spininterpretation}%
  \BibitemOpen
  \bibinfo {note} {It is also possible to interpret {$\hat{C}_k$} as the total
  {$\hat{S}_z$} of two spin-1/2 systems, where {$|01\rangle \pm |10\rangle$}
  correspond to {$|\pull\uparrow \uparrow \rangle$} and {$|\pull\downarrow
  \downarrow \rangle$}, and {$|00\rangle$} and {$|11\rangle$} correspond to
  {$|\pull\uparrow \uparrow \rangle \pm |\pull\downarrow \downarrow \rangle$}.
  Then $\smash{\hat{C}}$ represents the total magnetization in the
  lattice.}\BibitemShut {Stop}%
\end{thebibliography}

%

\end{document}